\documentclass[11pt,a4paper]{article}
\usepackage{jheppub}
\usepackage{multirow}
\usepackage{subfigure}
\title{Aspects of approximation in modelling of the symmetry energy.}
\author{Ilona Bednarek,}
\author{Monika Pienkos,}
\author{Jan Sladkowski}
\author{and Jacek Syska}
\affiliation{ Institute of Physics\\
University of Silesia, ul. 75 Pu\l ku Piechoty 1, 41-500 Chorz\'ow, Poland}
\emailAdd{ilona.bednarek@us.edu.pl}
\emailAdd{monika.pienkos@us.edu.pl}
\emailAdd{jan.sladkowski@us.edu.pl}
\emailAdd{jacek.syska@us.edu.pl}
\abstract{Quality of approximations is an important issue  in modelling nuclear matter. It is shown that the Pad\' e approximation provides  a useful tool  for describing the symmetry energy in highly asymmetric systems. The  focus is on the symmetry energy constraints.  Some implications of the obtained results for astrophysics are also discussed.}
\notoc
\begin{document}
\maketitle
\section{Introduction}

The correct form of the nuclear matter equation of state (EoS) is of much importance not only for understanding the
structure and properties of finite nuclei but   has also far-reaching
consequences for neutron stars. There are many theoretical models that describe their structure and properties in a satisfactory manner. A variety of these models results from the lack of detailed observational data that could impose adequate constraints on the  EoS describing dense asymmetric nuclear matter. This fact justifies the desirability of constructing more sophisticated theoretical models that admit reconstruction of not only neutron star parameters  but also would allow  the correct interpretation of the  experimental data involving finite nuclei and especially heavy ion collisions (HIC). The latter is a type of experiments  particularly promising because it offers the possibility to explore an EoS of nuclear matter, that encodes information about the symmetry energy, under conditions that are changed in a controlled manner.
Depending on the energy of the beam and the isotopic composition  it is possible to reproduce and study  nuclear matter for various range of density and neutron-proton asymmetry.
In such aspect, within the limits that can be achieved for a given experiment and for well-defined isospin sensitive observables, HIC is the unique tool to gain information on asymmetric nuclear matter. However, the key  fact that affects the interpretation of the obtained  experimental data is that HICs are non-equilibrium processes and every data analysis involve a transport model.
Thus, at the basis of the experimental determination of symmetry energy lies the selection of the appropriate observables and the theoretical model that is involved in their description.
HICs in the energy range from Fermi energies
 up to energies of several GeV per particle, through appropriate selection of phenomena   enable the analysis of the form of symmetry energy in various density range. The dense, hot and isospin asymmetric nuclear matter created during HIC is subject to evolution, which involves expansion and cooling thus,  creating properly diversified physical conditions for extracting information on the symmetry energy.
The measurements of individual quantities depend on the energy that is achieved during  experiments. To define observables sensitive to the symmetry energy it is important to correctly describe the produced clusters and fragments  in transport models involved. Considering some of the selected key observables in HIC that allow to gather information on the symmetry energy dependence on density the following examples can be quoted.
At Fermi energy (below and around the saturation density $n_{0}$) fragmentation, isospin fractionation and isospin transport are the key factors that determine the isospin contents of the reaction products.
At intermediate energies during the initial phase of the collision the pre-equilibrium emission of high energy particles and light fragments take place.  The measured observables, which are  considered as  strong  probe of the symmetry energy are the neutron and proton ratio  $R(n/p)$ and at higher energies $R(t/{}^{3}$He) \cite{2006:Famiano, 2008:Zhang, 2017:Pfabe}.
 Analysis of data supports the conclusion that the magnitude of $R(n/p)$ depends on the stiffness of the symmetry energy and that the   impact of the symmetry energy increases  with the increasing value of the neutron-proton asymmetry.
Cancelation of  residual effects that are not related to the symmetry energy demands measurements  of the double $n/p$ ratio
\begin{equation}
DR(n/p)=\frac{R_{{}^{124}Sn+{}^{124}Sn}(n/p)}{R_{{}^{112}Sn+{}^{112}Sn}(n/p)}
\end{equation}
for two different systems.  By comparing  double ratios $DR(n/p)$ with the predictions of the ImQMD transport model one can get constraints on the symmetry energy.
In particular, this allows  inference in the  problem of the momentum dependence of the symmetry mean-field potential that leads to differences between neutron and proton effective masses \cite{2014:Zhang, 2016:Coupland}.  The analysis of the
  correlated emission of generated particles was performed in order to evaluate and quantify the symmetry energy in the high density limit. In general, the phenomenological description of the  collective expansion is called the flow. In a non-central collisions the anisotropic flow concerns the anisotropy in particle momentum distributions correlated with the reaction plane. Information that can be extracted from observation of anisotropic flow bases on the analysis that is performed in terms of the Fourier series expansion of the azimuthal distribution \cite{2009:Bartke, 2011:Snellings, 2017:Pfabe}
\begin{equation}
\label{eq:el_flow}
E\frac{d^3N}{dp}=\frac{1}{2\pi}\frac{d^{2}N}{p_{t}dp_{t}dy}\left( 1+2\sum_{n=1}^{\infty}v_{n}(p_{T},y)\cos(n(\theta-\psi_{RP}))\right),
\end{equation}
where $E$ and $p$ denotes the energy and momentum of a particle, $p_{t}$ is  the transverse momentum,  $\theta$  the  azimuthal angle for the fixed rapidity $y$,  $\psi_{RP}$ defines the position of the reaction plane. For the isotropic emission $v_{n}=0$ for all $n$, $v_{n}\neq 0 $ indicates anisotropic emission with the
$v_{1}$ and $v_{2}$ coefficients that allows for the characteristics of the directed and elliptic flow, respectively. The analysis of the recently measured data by the ASY-EOS Collaboration \cite{2016:Russotto}, for the reaction ${}^{197}Au+{}^{197}Au$ leads to the
estimation of the power-law coefficient describing the density dependent potential part of the symmetry energy.

Phenomena controlled by the symmetry energy are related to the general idea of the isospin transport (difference between neutron $j_{n}$ and proton $j_{p}$ currents), which  results from the combined effects  of isospin and density gradients \cite{2017:Pfabe, 2005:Baran}
\begin{equation}
j_{n}-j_{p}\approx E_{sym}(n_b)\nabla I+\frac{\partial}{\partial n_{b}}I\nabla n_{b}.
\end{equation}
The isospin gradient with the coefficient that is proportional to the symmetry energy determines diffusion  and the density gradient defined  as drift is proportional to the slope of the symmetry energy. The observable that bases on the isospin diffusion and gives information on the symmetry energy is the isospin equilibration.
Isospin
equilibrium is one of several types of equilibrium that
excited nuclear matter can achieve. Its theoretical description requires modelling time evolution of the system and therefore involves transport model dependence. If such large spacetime volume equilibrium is actually achieved in experiments can be disputed.
 In all discussed energy regimes we encounter  non-equilibrium dynamical processes, which have to be modelled (transport theory). Although this approach has quite correct theoretical description,
various approximations  are inevitable what  makes things even more complicated and produces additional uncertainties. To deal with this problem an approach based on Pad{\'e} approximation is put forward. \\

The paper is organized as follows. Basic concepts related to the description of the symmetry energy are presented in Section II. The Pad\'{e} approximation approach is described in Section III. Then, in Section IV, experimental limitation concerning high density limit are discussed with the  emphasis on the role of the forth order term in the symmetry energy. Details of the theoretical models relevant for the analysis of asymmetric nuclear matter and neutron star matter are given in Section V. Main results are shown in Sections VI and VII.
\section{Symmetry energy}
The energy density of  asymmetric nuclear matter, in the most general form, can be formulated  in terms of the baryon density $n_{b} = n_{n}+n_{p}$ and isospin asymmetry parameter $\delta_{a}$, which  determines the difference between concentrations of neutrons $n_n$ and protons $n_{p}$ and can also be expressed through the relative proton fraction $Y_{p}$:
\begin{equation}
\delta_{a}=\frac{n_{n}-n_{p}}{n_{b}}=1-2Y_{p},\qquad Y_{p}=\frac{n_p}{n_b}.
\end{equation}
States with different neutron and proton concentrations are characterised by different energies. These energy differences significantly figure in the symmetry energy, which   for the system that comprises  $N$ nucleons ($N=N_{n}+N_{p}$, $n_{b}=N/V$, $n_{i}=N_{i}/V \,\, i=n,p$), can be written as
\begin{equation}
\label{eq:decomp}
\epsilon_{sym}(n_{b},\delta_{a}) = \epsilon(n_{b},\delta_{a})-\epsilon(n_{b},0).
\end{equation}
Equation (\ref{eq:decomp}) shows the decomposition of  the energy  $\epsilon(n_{b},\delta_{a})$ into the symmetric part $\epsilon(n_{b},0)$ with equal numbers of protons and neutrons  and the isospin dependent  one.
A commonly used method that allows one to separate the isospin dependent part of the energy and also offers the possibility to perform more detailed analysis of the symmetry energy is a Taylor expansion of the function $\epsilon(n_{b},\delta_{a})$, which after introducing a compact notation:
$\mathbf{x}=(n_{b},\delta_{a})$ and $\mathbf{x_{0}}=(n_{0},0)$   can be written in the following way:
\begin{equation}
\label{eq:Taylor1}
\epsilon(\mathbf{x})=\epsilon(\mathbf{x_{0}})+[(\mathbf{x}-\mathbf{x_{0}})
\cdot\mathbf{\nabla}
\epsilon(\mathbf{x_{0}})] +\frac{1}{2}[(\mathbf{x}-\mathbf{x_{0}})\cdot(H(\mathbf{x})\cdot(\mathbf{x}-
\mathbf{x_{0}}))]+\ldots,
\end{equation}
where $H(\mathbf{x})$ is the matrix of second derivatives of the function $\epsilon(\mathbf{x})$ (Hessian matrix).
Such decomposition is a proven method of analysing  properties of nuclear matter. It has been shown that the
Taylor series of  the function $\epsilon(n_{b},\delta_{a})$  provides an approximation of the energy density in some open neighbourhood around $(n_{b},\delta_{a})= (n_{0},0)$.
 The truncated series which includes only second order terms  represents the so-called parabolic approximation
 is widely used to describe properties of nuclear matter in a limited
 range of   density (near the
nuclear saturation density $n_{0}$) and neutron-proton asymmetry (nuclear matter akin to the symmetric one $\delta_{a}\sim 0$).
The approximation of the function $\epsilon(n_{b},\delta_{a})$ that  refers to the Taylor expansion  allows one to specify definition of the symmetry energy.
Considering the Taylor  expansion of the energy density $\epsilon(n_b,\delta_{a})$ in terms of neutron-proton asymmetry
$\delta_{a}$, the relation that allows one to obtain the conventional definition of the symmetry energy can be derived:
\\
\\
\begin{equation}
\label{eq:Taylor}
\epsilon(n_{b},\delta_{a})=\epsilon(n_{b},0)+\frac{1}{2}\frac{\partial^{2}\epsilon(n_{b},\delta_{a})}{\partial\delta_{a}^{2}}\bigg{|}_{\delta_{a}=0}\delta_{a}^{2}
+\frac{1}{24}\frac{\partial^{4}\epsilon(n_{b},\delta_{a})}{\partial\delta_{a}^{4}}\bigg{|}_{\delta_{a}=0}\delta_{a}^{4}+\ldots
\end{equation}
The above expression includes terms up to the forth order in the isospin asymmetry parameter. Terms with odd powers in $\delta_{a}$ vanish. Introducing the notation:
\begin{eqnarray}
\epsilon(n_{b},0)&=&\epsilon_{0}(n_{b}), \\ \nonumber
 E_{2,sym}(n_{b})&=&\frac{1}{2}\frac{\partial^{2}\epsilon(n_{b},\delta_{a})}{\partial\delta_{a}^{2}}\bigg{|}_{\delta_{a}=0},\\ \nonumber
E_{4,sym}(n_{b})&=&\frac{1}{24}\frac{\partial^{4}\epsilon(n_{b},\delta_{a})}{\partial\delta_{a}^{4}}\bigg{|}_{\delta_{a}=0}
\end{eqnarray}
the energy density $\epsilon(n_{b},\delta_{a})$ can be rewritten as follows:
\begin{equation}
\label{eq:Taylor2}
\epsilon(n_{b},\delta_{a})=\epsilon(n_{b},0)+E_{2, sym}(n_{b})\delta_{a}^2+E_{4,sym}(n_b)\delta_{a}^{4}+\ldots \, ,
\end{equation}
where  $E_{2,sym}$ and $E_{4,sym}$ are density dependent. Currently, there is no knowledge about the density dependence of the function $E_{4,sym}(n_{b})$.
 The leading order coefficient in series (\ref{eq:Taylor2}) proportional to $\delta_{a}^{2}$ is  widely accepted as a definition of the symmetry energy.
Subsequently  $\epsilon(n_b,\delta_{a}=0)$ and $E_{2,sym}(n_{b})$ can be expanded in a Taylor series  around the equilibrium density $n_0$:
\begin{equation}
\varepsilon(n_b,\delta_{a}=0)=\varepsilon(n_0)+
\frac{1}{2!}\frac{\partial^2\varepsilon}{\partial
n_b^2}\bigg{|}_{n_{0}}u^2+\ldots \label{en0}
\end{equation}
\begin{equation}
E_{2,sym}(n_b)=E_{2,sym}(n_0)+\frac{\partial E_{2,sym}}{\partial
n_b}\bigg{|}_{n_{0}}u+\frac{1}{2!}\frac{\partial^2E_{2,sym}}{\partial
n_b^2}\bigg{|}_{n_{0}}u^2+\ldots,\label{esym}
\end{equation}
where $u=(n_{b}-n_{0})/3n_{0}$. Equations ( \ref{eq:Taylor2}, \ref{en0}, \ref{esym}) allows one to express the nuclear matter EoS through a series of coefficients:
\begin{equation}
\label{eos:coef}
\epsilon(n_b,\delta_{a})=\epsilon(n_0)+E_{sym}(n_{0})\,\delta_{a}^{2}+ L\,\delta_{a}^{2}\,u+
(K_0+K_{sym}\,\delta_{a}^{2})\,u^2,
\end{equation}
where $\epsilon(n_{0})$ is the
binding energy of symmetric nuclear matter and  $K_{0}$ is the incompressibility.
Parameters that  characterize the isospin dependent part of the EoS are: the symmetry energy coefficient $E_{2,sym}(n_{0})$, its density slope $L$ and curvature $K_{sym}$. The last two parameters are given by the following equations:
\begin{equation}
L(n_{0})=3n_{0}\frac{dE_{sym}(n_{b})}{dn_{b}}\bigg{|}_{n_{b}=n_{0}},
\end{equation}
\begin{equation}
K_{sym}(n_{0})=9n_{0}^{2}\frac{d^{2}E_{sym}(n_{b})}{dn_{b}^{2}}\bigg{|}_{n_{b}=n_{0}}.
\end{equation}
Given data from many terrestrial experiments and using constraints coming from astrophysical observations it is possible to   estimate consistently the values of the symmetry energy coefficient   $E_{2,sym}(n_{0})= 31.6 \pm 2.66 $ MeV and the slope $L(n_{0})=58.9\pm 16$ MeV \cite{2017:Li}.

\subsection{Pad{\'e} approximation}

The need to solve various
modelling and computation problems have resulted in the development of approximation theory that provided us with such tools as
the Pad\'{e} approximants
\cite{Encyclopedic}.
The first computational problem in the analysis of the symmetry energy
by considering the Taylor series of the energy density $\epsilon(n_{b},\delta_{a})$ of the nuclear matter expanded around the isospin asymmetry parameter $\delta_{a}$,
consists in the implicit-like definition of the symmetry energy. The second one arises while trying to find possibly the best analytic formula for
values of the symmetry energy in a region, where experimental data are not available, i.e., at very high densities.
Thus, although
the
definition of the symmetry energy could
be accepted
as the coefficient in the second order term in
the
Taylor series for the energy density,
yet in
the case of high densities that are relevant for neutron star interiors such expansion
can be questioned \cite{high-densities-problem}, \cite{2017:Gonzales}.
Now, a Pad\'{e} approximant
gives better approximation of the initial function than the particular truncated Taylor series.
Also the Pad{\'e} approximants may  still be used in that parts of the function domain,
where the Taylor series does not converge, i.e., beyond
its disc of convergence \cite{Suetin}. \\
%
%
%
The Pad{\'e} approximants can be viewed as formal Gaussian quadrature methods \cite{Brezinski}.
Below are recalled some basic
results on the Pad{\'e} approximants.
Let $c(z)$ be a formal power series
$c(z) \equiv \sum_{i=0}^{\infty} c_{i} z^{i}$ of the analysed function $f(z)$.
For any pair of nonnegative integers $(L, M)$, the $[L, M]$-th Pad{\'e} approximation
of $f(z)$ is defined as a rational function
$P^{L}_{M}  \equiv \frac{A_{L}}{B_{M}} = \frac{\sum_{i=0}^{L} a_{i} z^{i}}{\sum_{i=0}^{M} b_{i} z^{i}}$, satisfying the condition that all terms in the formal power series
$(\sum_{i=0}^{M} b_{i} z^{i})(\sum_{i=0}^{\infty} c_{i} z^{i})
- \sum_{i=0}^{L} a_{i} z^{i}$ should vanish up to the term $z^{L+M}$.
Provided that every Hankel determinant is different from zero, the Pad{\'e} approximation is uniquely determined \cite{Encyclopedic}.
%
%
Then, $b_{0}$ can be chosen uniquely as equal to 1. \\
%
%
It is essential to know
the truncation error,
$R_{L,M}(z) = f(z) - P^{L}_{M}$,
which arises when the function $f(z)$ is replaced by its approximation.
%
%
%
Now, given $L_{k}+M_{k} \rightarrow \infty$ as $k \rightarrow \infty$, the limit $R_{L_{k},M_{k}}(z) \rightarrow 0$ for $L_{k}=M_{k}$, which  follows from  the
Pad{\'e} conjecture due to Baker, Gammel, and Wills  \cite{Pade-Approximants}.
The assumption of  convergence of the diagonal ($L=M$)
sequence
of approximants \cite{Pade-Approximants, 1996:Baker, Stahl,theoret-vs-numer-convergence}
is
the foundation of many calculations and applications \cite{applications-1, applications-2, 2018:Grober, 1952:Yang, 1999:Nickel, 1985:Kubo, 1990:Berretti, applications-3, 1997:Stahl, 1997:Gilewicz, 2009:Bessis, applications-4, 1994:Llave, 1995:Berretti, 2001:Berretti}.
Calculations in this paper also are  based  on its validity.
\\
The Pad{\'e} approximant $P^{2}_{2}$ is used  in the analysis of the energy density $\epsilon(n_{b},\delta_{a})$ of asymmetric nuclear matter which was performed in this paper.
%
%
As the Maclaurin series of
$\epsilon(n_{b},\delta_{a})$ taken with respect to the asymmetry
parameter $z \equiv \delta_{a}$ has the vanishing odd coefficients thus, $P^{2}_{2}$ is the lowest, nontrivial Pad{\'e} approximant. As
$c_{1}=c_{3}=0$
\begin{eqnarray}
\label{Pade 2 2 Nieparzyste zero}
P^{2}_{2}(n_{b}, \delta_{a})
= \frac{c_{0} c_{2} +  (c_{2}^{2} - c_{0} c_{4}) \delta_{a}^{2}}{c_{2} - c_{4} \delta_{a}^{2}}
= c_{0} + \frac{c_{2}^{2}}{c_{2} - c_{4}\delta_{a}^{2}}\delta_{a}^{2} \; ,
\end{eqnarray}
where
the coefficients $c_{i}$, $i=0,2,4$, in the Maclaurin series are  functions of the baryon number density $n_{b}$ and are equal to  the corresponding coefficients in Taylor expansion:
$c_{0}(n_{b})\equiv \epsilon (n_{0},0)$,
$c_{2}(n_{b}) \equiv E_{2,sym}(n_{b})$
and $c_{4}(n_{b}) \equiv E_{4,sym}(n_{b})$.
Finally, there are several algorithms that relate Taylor series and Pad\'{e} approximants and it is possible to compare  the
effectiveness of both approaches \cite{algorithms}.
%
%
%

Basing on the Pad{\'e} approach and using the standard definition of the  symmetry energy as the coefficient of  $\delta_{a}^{2}$ term, the following formula for the symmetry energy is obtained
\begin{equation}
\label{eq:esym_Pade}
E_{sym}^{P}(n_{b},\delta_{a})=g(n_b)\left(\frac{1}{1-f(n_{b})\delta_{a}^{2}}\right),
\end{equation}
where $g(n_{b})=E_{2,sym}(n_{b})$ and $f(n_{b})=E_{4,sym}(n_{b})/E_{2,sym}(n_{b})$. The symmetry energy given by Eq.(\ref{eq:esym_Pade}) is a function that depends not only on density but also on isospin asymmetry.
The introduction of the function $f(n_{b})$, which describes the ratio of the quartic and quadratic terms  has raised the question about the role and importance of the quartic term $E_{4,sym}(n_b)$.
The form of the symmetry energy, which depends only on the density is recovered ($E_{sym}^{P}(n_{b},\delta_{a})= E_{2,sym}(n_{b})$),  in the case when the quartic term  is neglected, $E_{4,sym}(n_{b}) = 0$.
Assuming self-similar density dependence of $E_{2,sym}(n_{b})$ and $E_{4,sym}(n_{b})$ functions, which thereby preserve the same density profiles, the function $f(n_{b})$ can be replaced by a constant parameter $q$. This particular case is analysed in Section V.
\subsection{The importance of the 4-th order term}
Symmetry energy definition formulated on the basis of the parabolic approximation may be justified in case of finite nuclei, where the value of the isospin asymmetry parameter $\delta_{a}$ is low. However, it seems to be inappropriate  when it  increases significantly reaching  a value close to one ($\delta_{a}\rightarrow 1$) in the extreme case.
Unfortunately, there are no experimental constraints on  $E_{4,sym}(n_{b})$ but, yet  there are several theoretical estimations of the value of  the 4-th order term at saturation density $E_{4,sym}(n_{0})$ which are collected in Table \ref{tab:S4_theory}.
\begin{table}
\caption{\label{tab:S4_theory}The value of the $E_{4,sym}(n_{0})$ obtained on the basis of different theoretical models}
\vspace{2mm}
\begin{center}
\begin{tabular}{l|l}
\hline
Theoretical model & Value of $E_{4,sym}(n_{0})$ [MeV]\\
\hline
Interacting Fermi gas model (\citep{2015:Cai})& $7.18 \pm 2.52$ \\ \hline
Chiral pion-nucleon dynamics (\citep{2015:Kaiser})& $1.5$ \\  \hline
Quantum molecular dynamics (QDM) (\citep{2016:Nandi})&$3.27\sim 12.7$\\  \hline
Relativistic mean field models (\citep{2012:Cai})& $\leq 2$\\  \hline
Extended nuclear mass formula (\citep{2017:Wang})& $20 \pm 4.6$ \\  \hline
\end{tabular}
\end{center}
\end{table}
\section{Symmetry energy aspects of systematics}
Description of isospin asymmetric nuclear matter involves a variety of theoretical approaches. Among others, three main categories were developed:
  microscopic, phenomenological and effective field theory.
The first of them applies to methods that incorporate
realistic interactions  modelled on the high precision free space nucleon-nucleon (NN) scattering data and the deuteron properties. The constructed NN interactions have to be combined with many-body calculations of nuclear systems. The basic approaches to the solution of  many-body problems are of the   variational-type  and of Brueckner-type \cite{1988:Wiringa}.
However, in the case of realistic NN interaction models there is a problem in reproducing  the saturation point.
The popular versions of the Argonne AV14 or
Urbana UV14 potentials used in the variational calculations did not reproduce
satisfactory saturation properties of nuclear matter \cite{1988:Wiringa}. The
phenomenological three-nucleon interaction with parameters
adjusted to obtain, together with the two-nucleon interaction, the
correct value of the nuclear matter saturation properties, has to
be introduced \cite{2002:Zuo}.
The non-relativistic case of the Brueckner-type
calculations, the Brueckner-Hartree-Fock (BHF) \cite{1991:Bombaci}
method which incorporates the realistic two-nucleon potentials
(for example the Bonn or Paris potential) also  does not saturate at
empirical density.

The relativistic Dirac Brueckner-Hartree-Fock
(DBHF) approach provides a better description of the nuclear
saturation properties of symmetric nuclear matter
\cite{2009:Gogelein}. In this case the saturation point  of nuclear matter is shifted to its empirical value due to the density dependence of the nucleon spinor and  consequently the medium dependence of the NN interaction \cite{2017:Muther}.
The second category of theoretical approaches, phenomenological models of nuclear matter EoS,  bases on the effective density dependent interactions, characterised by  parameters that  have been  determined from experimental data. Calculations performed in Hartree-Fock and Thomas-Fermi approximations for the most popular effective density dependent interactions, the Skyrme or Gogny forces,
satisfactorily reproduce various properties of nuclei and nuclear matter. Another phenomenological approach are relativistic mean field (RMF) models in which, in case of two component system, neutrons and protons are main constituents and description of strong interaction bases on the exchange of different kinds of mesons.
\cite{2018:Typel}. There are different extensions to this approach, which among themselves create separate classes of models. One of the models introduces nonlinear self and mixed meson couplings  in order to properly include medium effects in the description of finite nuclei and nuclear matter.  An alternative group of models are those that contain density-dependent meson-nucleon coupling constants.

The chiral effective field theory (EFT) has been also successfully applied to the modelling of NN interactions giving consistent description of two- and many-nucleon forces.
\subsection{Experimental constraints}
In general, knowledge about current theoretical and experimental limitations on the form of the symmetry energy in the full density range, below and above the saturation density, is very incomplete.
In this section a very short summary of the most relevant experimental and theoretical results have been collected  in order to present a certain minimum of information which seems to be indispensable to  deal with the problem of the high density dependence of the symmetry energy. In studies on this subject a very important issue is the selection of  appropriate observables  carrying,  for a given density range,  information about the symmetry energy which  is not  measured directly in experiments.
Results that have been obtained depend ultimately on  the range of studied density.
Isoscaling analysis of the yields of nuclei with $A \leq 4$ reveals that at very low density ($n_{b}\ll n_{0}$) the symmetry energy calculated on the basis of virial expansion \cite{2006:Horowitz} exceeds the one resulting from the mean field calculations \cite{2007:Kowalski}. Quantum statistical approach that includes the formation of clusters in nuclear matter follows this experimental results \cite{2010:Natowitz}.
Experiments, carried out in terrestrial  laboratories, dedicated to analysis of  the symmetry energy below and in close vicinity of the saturation point uses  among others  the following isospin-sensitive observables: atomic masses, the heavy ion multifragmentation reactions, measurements of neutron skin in antiprotonic atoms, Pygmy Dipole Resonance in ${}^{208}$Pb, Giant Dipole Resonance in ${}^{208}$Pb,
 neutron skins of heavy nuclei, isospin diffusion in heavy ion reactions, excitation energies of isobaric analog states, isoscaling of fragments from intermediate energy heavy ion collisions, electric dipol polarizability in the Pygmy Dipole Resonance \cite{2014:Horowitz}.
Results obtained for the existing experimental data and on the basis of theoretical models  allow for the analysis of the  symmetry energy at  densities $n_{b}\leq n_{0}$
  and for drawing coherent conclusions on the behaviour of the symmetry energy for sub-saturation densities.

 Unfortunately, the symmetry energy
remains largely unconstrained both theoretically and experimentally at densities above saturation density $n_{0}$ and   this problem is far from being settled.  Most of  information on this subject comes from  HICs data.
It is predicted that this data  can be related to isospin sensitive observables that are associated with the early phases of collisions. Various such observables have been  specified.
However, the ongoing processes in HICs are non-equilibrium ones and their interpretation involve transport models.
This, despite the good accuracy of the experimental data,
leads to a lack of consistency between some of the results.
In such cases, contradictory constraints on the density dependence of the symmetry energy can be obtained.
In this context, the task of finding the most adequate isospin sensitive observables becomes very important.
A very promising one is the elliptic flow ratio of neutrons with respect to protons and  light charged complex particles in reactions of neutron-rich systems at pre-relativistic energies. It is given by the  elliptic flow parameters  difference $v_2^{n}-v_2^{p}$ and ratio  $v_{2}^{n}/v_{2}^{p,H}$ (\ref{eq:el_flow}).

In many cases analysis of the symmetry energy is based on its density dependence expressed in the form of a power law:
\begin{equation}
\label{eq:power_law}
E_{sym}(n_{b})=E_{sym}^{pot}(n_{b})+E_{sym}^{kin}(n_{b})=  22 MeV \left(\frac{n_{b}}{n_{0}}\right)^{\gamma}+ 12 MeV \left(\frac{n_{b}}{n_{0}}\right)^{2/3}.
\end{equation}
Equation (\ref{eq:power_law}) includes the kinetic and potential terms with $\gamma$ being a parameter  that completely characterizes the potential  part of the symmetry energy.
The existing early data from the FOPI-LAND experiment \cite{1993:Leifels, 1994:Lambrecht} have been compared with calculations performed with the ultra-relativistic quantum molecular dynamics (UrQMD) transport model \cite{2005:Li, 2006:Li, 2006a:Li} and with the power-low parametrization of the density dependence of the potential parts of the symmetry energy.
The obtained result  allows to restrict the density dependence of the symmetry energy what corresponds to the following value of the coefficient $\gamma = 0.9\pm 0.4$.
In the central zone of HICs the value of density of the order of $2 \times n_{0}$ can be achieved at moderate energy of 400 MeV/nucleon \cite{2016:Russotto}.
In order to improve the accuracy of the experimental flow parameters   a new ASY-EOS experiment  \cite{2016:Russotto} was carried out on ${}^{197}$Au +${}^{197}$Au, ${}^{96}$Zr +${}^{96}$Zr and ${}^{96}$Ru +${}^{96}$Ru collisions at 400 MeV/nucleon.
The performed analysis, with particular effort directed at the improvement of the statistical accuracy leads to the  power law coefficient  $\gamma = 0.72\pm 0.19$ \cite{2016:Russotto}.


\section{The model}
The theoretical model on the basis of which the calculations were carried out
 comprises nucleons interacting through the exchange of the scalar meson $\sigma$ and vector mesons $\omega$ and $\rho$.
The Lagrangian density can be given as a sum ${\mathcal{L}}={\mathcal{L}_{0}}+{\mathcal{L}_{M}}+{\mathcal{L}_{int}}$,  where the
${\mathcal{L}}_{0}$ denotes the Lagrangian of free nucleons
\begin{equation}
\mathcal{L}_{0}=\bar{\psi}_{N}\left(i\gamma^{\mu}\partial_{\mu}-m_{N}\right)\psi_{N},
\end{equation}
and $$\psi_{N}=\left(\begin{array}{c}
\psi_{p}\\
\psi_{n}
\end{array}\right)$$ with neutron and proton spinor  $\psi_n$ and $\psi_p$.
${\mathcal{L}_{M}}$ is the Lagrangian of  free meson fields $\sigma, \omega$ and $\rho$:
\begin{equation}
{\mathcal{L}_{M}}=
\frac{1}{2}\partial_{\mu}\sigma\partial^{\mu}\sigma
-\frac{1}{2}m_{\sigma}^{2}\sigma^{2}+\frac{1}{2}m_{\omega}^{2}(\omega_{\mu}\omega^{\mu})+
\frac{1}{2}m_{\rho}^{2}(\rho_{\mu}^{a}\rho^{a\mu})
-\frac{1}{4}\Omega_{\mu\nu}\Omega^{\mu\nu}
-\frac{1}{4}\mathbf{R}_{\mu\nu}\mathbf{R}^{\mu\nu},
\end{equation}
where $\Omega_{\mu\nu}=\partial_{\mu}\omega_{\nu}-\partial_{\nu}\omega_{\mu}$ and $\mathbf{R}^{\mu\nu}=\partial_{\mu}\rho_{\nu}^{a}-\partial_{\nu}\rho_{\mu}^{a}$ are the field tensors.
The interaction Lagrangian ${\mathcal{L}_{int}}$ has the form
\begin{equation}
{\mathcal{L}_{int}}=g_{\sigma}\sigma\bar{\psi}_{N}\psi_{N}- g_{\omega }\omega_{\mu}\bar{\psi}_{N}\gamma^{\mu}\psi_{N}-\frac{1}{2}g_{\rho}\rho_{\mu}^ {a}\bar{\psi}_{N}\gamma^{\mu}\bar{\tau}\psi_{N}  -U(\sigma,\omega,\rho).
\end{equation}
The potential $U_{eff}(\sigma,\omega,\rho)$ includes nonlinear scalar  and vector
meson terms:
\begin{eqnarray}
U_{eff}(\sigma,\omega, \rho) &=&\frac{1}{3}g_{2}\sigma^{3}+\frac{1}{4}g_{3}\sigma^{4}- \frac{1}{4}c_{3}(\omega_{\mu}\omega^{\mu})^{2} \\ \nonumber
&-&\left(\Lambda_{S}(g_{\sigma}g_{\rho})^{2}(\omega_{\mu}\omega^{\mu})\sigma^{2}+\Lambda_{V}(g_{\omega}g_{\rho})^{2}(\omega_{\mu}\omega^{\mu})\right)(\rho_{\mu}^{a}\rho^{a\mu})
\end{eqnarray}
and its form is dictated by the need for correct description of both the bulk properties of finite nuclei and nuclear matter. \\
The mixed isoscalar - isovector coupling $\Lambda_{V}(g_{\omega}g_{\rho})^{2}(\omega_{\mu}\omega^{\mu})(\rho_{\mu}^{a}\rho^{a \mu})$ \cite{2001:Horowitz, 2009:Sharma, 2015:Bednarek} enhances the isovector sector of the model and thereby the ability of  the   density dependence of the symmetry energy to be modified.
The considered parametrisation (Table \ref{tab:TM1}) gives satisfactory results  when describing the properties of finite and infinite nuclear systems.
However, the original model may not be sufficient to correctly describe the form of the symmetry energy and especially its high density limit.
The nonlinear mixed isoscalar-isovector couplings are particularly important as their function is to undertake the modification of the density dependence of the symmetry energy \cite{2014:Bednarek}.
\begin{table}
\caption{The TM1 parameter set \cite{1994:Sugahara}}
\begin{center}
\begin{tabular}{l|l|l|l|l|l}
\hline
         scalar field $\sigma$& $m_{\sigma} = 511.198$ MeV  & $g_{\sigma}=10.029$ & $g_{2}=$7.2327fm$^{-1}$ & $g_{3}=0.6183$ &-\\ \hline
vector field $\omega$     & $m_{\omega} =783$ MeV& $g_{\omega}=12.614$ & $c_{3} = 71.308$ & -&- \\ \hline
\multirow{2}{*}{}vector field $\rho$ & \multirow{2}{*}{}$m_{\rho} =770 $ MeV & $\Lambda_{V}$ &  0.0&0.015  &0.03\\ \cline{3-6}
                  &                  & $g_{\rho}(\Lambda_{V})$ & 9.2644  & 10.6767&11.0964 \\
                  \hline
\end{tabular}
\end{center}
\label{tab:TM1}
\end{table}
The Lagrangian function ${\mathcal{L}}={\mathcal{L}_{0}}+{\mathcal{L}_{M}}+{\mathcal{L}_{int}}$ leads to  field equations that were solved in the mean field approximation, which  requires substitution of the meson fields operators by their expectation values: $\sigma(x)\rightarrow <\sigma(x)>\equiv s_{0}$, $\omega^{\mu}(x)\rightarrow <\omega^{\mu}(x)\delta_{0\mu}>\equiv w_{0}$, $\rho^{\mu a}(x)\rightarrow <\rho^{\mu a}(x)\delta_{0\mu}\delta^{3a}>\equiv r_{0}$. Adopting  this approach,  the following form of the  equations of motion can be obtained:
\begin{equation}
m_{\sigma}^{2}s_{0} =
-g_{2}s_{0}^{2}-g_{3}s_{0}^{3}+\Lambda_{S}(g_{\sigma}g_{\rho})^{2}s_{0}r_{0}^{2}+g_{\sigma}n_{s},
\label{eqm1}
\end{equation}
\begin{equation}
m_{\omega}^{2}w_{0} + c_{3}w_{0}^{3}+\Lambda_{V}(g_{\omega}g_{\rho})^{2}r_{0}^2w_{0} = g_{\omega}n_{b},
\label{eqm3}
\end{equation}
\begin{equation}
m_{\rho}^{2}r_{0}+(\Lambda_{V}(g_{\omega}g_{\rho})^{2}w_{0}^2+\Lambda_{S}(g_{\sigma}g_{\rho})^{2}s_{0}^2)r_{0} = g_{\rho}\delta_{a}n_{b},
\label{eqm4}
\end{equation}
\begin{equation}
(i\gamma_{\mu}\partial^{\mu}-m_{eff}(s_{0})-g_{\omega}\gamma^{0}w_{0}-g_{\rho}I_{3N}\gamma^{0}\tau^{3}r_{0})\psi_{N}=0.
\end{equation}
The scalar  $n_{s}$ and baryon $n_{b}$ densities  are given by
\begin{equation}
n_{s}=
\sum_{N=n,p}\frac{\gamma}{(2\pi)^3}\int_{0}^{k_{F,N}}\frac{m_{eff}(s_{0})d^3k}{\sqrt{(k^{2}+m_{eff}^{2}(s_{0}))}},
\end{equation}
$n_{b}=(\gamma/6\pi^{2})k_{F}^{3}$, where $\gamma$ is the spin-degeneracy factor and  $m_{eff}=m_{N}-g_{\sigma}s_{0}$ \cite{2001:Manka}.
Solutions of the field equations allow for the calculation of the energy density $\epsilon $ and pressure $P$ of nuclear matter. From the energy-momentum tensor the energy density $\epsilon = <T^{00}>$ and pressure $P=1/3\sum_{i}<T^{ii}>$ can be obtained and their forms in the  mean field limit are given by:
\begin{eqnarray}
\label{eq:energy}
\epsilon
&=&\frac{1}{2}m_{\omega}^{2}w_{0}^{2}
+\frac{1}{2}m_{\rho}^{2}r_{0}^{2}
+\frac{1}{2}m_{\sigma}^{2}s_{0}^{2}
+\frac{3}{4}c_{3}w_{0}^{4}+ 3\Lambda_{V}(g_{\omega}g_{\rho})^{2}w_{0}^{2}r_{0}^{2} \\ \nonumber
&+&\Lambda_{S}(g_{\sigma}g_{\rho})^{2}s_{0}^{2}r_{0}^{2}+\frac{1}{3}g_{2}\sigma^{3}+\frac{1}{4}g_{3}\sigma^{4}
+\sum_{N=n,p}\frac{2}{(2\pi)^{3}}\int_{0}^{k_{F,N}}d^3k\sqrt{k^{2}+m_{eff}^{2}(s_{0})},
\end{eqnarray}
\begin{eqnarray}
\label{eq:pressure}
 P
&=&\frac{1}{2}m_{\omega}^{2}w_{0}^{2}
+\frac{1}{2}m_{\rho}^{2}r_{0}^{2}
-\frac{1}{2}m_{\sigma}^{2}s_{0}^{2}+\frac{1}{4}c_{3}w_{0}^{4}+ \Lambda_{V}(g_{\omega}g_{\rho})^{2}w_{0}^{2}r_{0}^{2} \\ \nonumber
 &+&\Lambda_{S}(g_{\sigma}g_{\rho})^{2}s_{0}^{2}r_{0}^{2}-\frac{1}{3}g_{2}\sigma^{3}-\frac{1}{4}g_{3}\sigma^{4}
+\sum_{N=n,p}\frac{2}{3(2\pi)^{3}}\int_{0}^{k_{F,N}}d^3k\frac{k^{2}}{\sqrt{k^{2}+m_{eff}^{2}(s_{0})}}.
\end{eqnarray}
Using the parabolic approximation as the definition of the symmetry energy one can obtain  for the considered model its general form:
\begin{equation}
\label{eq:esym2}
E_{2,sym}(n_{b})=\frac{k_{F}^{2}}{6\sqrt{k_{F}^{2}+m_{eff}^{2}}}+\frac{g_{\rho}^{2}n_{b}}{2(m_{\rho}^{2}+\Lambda_{V}(g_{\rho}g_{\omega})^{2}w_{0}^{2}+\Lambda_{S}(g_{\rho}g_{\sigma})^{2}s_{0}^{2})}.
\end{equation}
Taking into regard corrections to the 4-th order in the Taylor expansion $E_{4,sym}(n_{b}) = E_{4,sym}^{kin}(n_{b}) +E_{4,sym}^{pot}(n_{b})$ the following additional contributions to the symmetry energy should be included:
\begin{equation}
\label{eq:esym4k}
E_{4,sym}^{kin}(n_{b})=\frac{k_{F}^{2}}{648}\frac{4m_{eff}^{4}+11m_{eff}^{2}k_{F}^{2}+10k_{F}^{4}}{(\sqrt{k_{F}^{2}+m_{eff}^{2}})^{5}}
\end{equation}
\begin{eqnarray}
\label{eq:esym4p}
E_{sym,4}^{pot}(n_{b})&=&\frac{g_{\rho}^{8}n_{b}^{3}}{2Q_{\rho}^{4}}+\left(\frac{\Lambda_{V}g_{\omega}^{4}w_{0}^{2}}{Q_{\omega}}-
\frac{\Lambda_{S}^{2}g_{\sigma}^{4}s_{0}^{2}}{Q_{\sigma}}\right)+\\ \nonumber
&+&\frac{g_{\sigma}^{2}n_{b}m_{eff}k_{F}^{2}}{24Q_{\sigma}(\sqrt{k_{F}^{2}+m_{eff}^{2}})^{3}}\left(\frac{4\Lambda_{S}g_{\sigma}g_{\rho}^{4}s_{0}n_{b}}{Q_{\rho}^{2}}-\frac{m_{eff}k_{F}^{2}}{3(\sqrt{k_{F}^{2}+m_{eff}^{2}})^{3}}\right),
\end{eqnarray}
where $Q_{\sigma}, Q_{\omega}$ and $Q_{\rho}$ are given by the following relations \cite{2012:Cai}
\begin{equation}
Q_{\sigma}=m_{\sigma}^{2}+g_{\sigma}^{2}\left(\frac{3n_{s}}{m_{eff}}-\frac{3n_{b}}{\sqrt{k_{F}^{2}+m_{eff}^{2}}}\right)+2g_{2}s_{0}+3g_{4}g_{3}s_{0}^{3},
\end{equation}
\begin{equation}
Q_{\omega}=m_{\omega}^{2}+3c_{3}(g_{\omega}w_{0})^{2}
\end{equation}
\begin{equation}
Q_{\rho}=m_{\rho}^{2}+\Lambda_{S}(g_{\sigma}g_{\rho})^{2}s_{0}^{2}+\Lambda_{V}(g_{\omega}g_{\rho})^{2}w_                                                                                                                                                                                                              {0}^{2}.
\end{equation}
The obtained relations allow one to calculate the symmetry energy  in the Pad{\'e} approximation.

\subsection{Matter of a neutron star}
The chemical composition of a neutron star is set with the assumption of charge neutrality and equilibrium  due to $\beta$ decay thus, the correct model of matter requires the presence of electrons and muons.
The Lagrangian function must be supplemented by the  Lagrangian of free leptons  $\mathcal{L}_{l}$:
\begin{equation}
\mathcal{L}_{l}=\sum_{l=e,\mu}\overline{\psi}_{l}(i\gamma^{\mu}\partial_{\mu}-m_{l})\psi_{l}.
\end{equation}
The appearance  of leptons gives additional contributions to the energy density $\epsilon_{l}$ and pressure $P_{l}$ of neutron star matter, where
\begin{equation}
\epsilon_{l}=\frac{2}{(2\pi)^{3}}\int_{0}^{k_{F}}d^3k\sqrt{k^{2}+m_{l}^{2}}
\end{equation}
and
\begin{equation}
P_{l}=\frac{2}{3(2\pi)^{3}}\int_{0}^{k_{F}}d^3k\frac{k^{2}}{\sqrt{k^{2}+m_{e}^{2}}}.
\end{equation}
The chemical equilibrium is characterised by correlations between chemical potentials, which in the process $p+e^{-}\leftrightarrow n+\nu_{e}$ leads to the following relation $\mu_{asym}=\mu_{n}-\mu_{p}=\mu_{e}-\mu_{\nu_{e}}$.  Assuming that neutrinos are no trapped in neutron star matter ($\mu_{\nu_{e}} = 0$), $\mu_{asym}$ reduces to $\mu_{asym}=\mu_{n}-\mu_{p}=\mu_{e}$.
\section{Results}
\begin{figure}
\centering \includegraphics[clip,width=9cm]{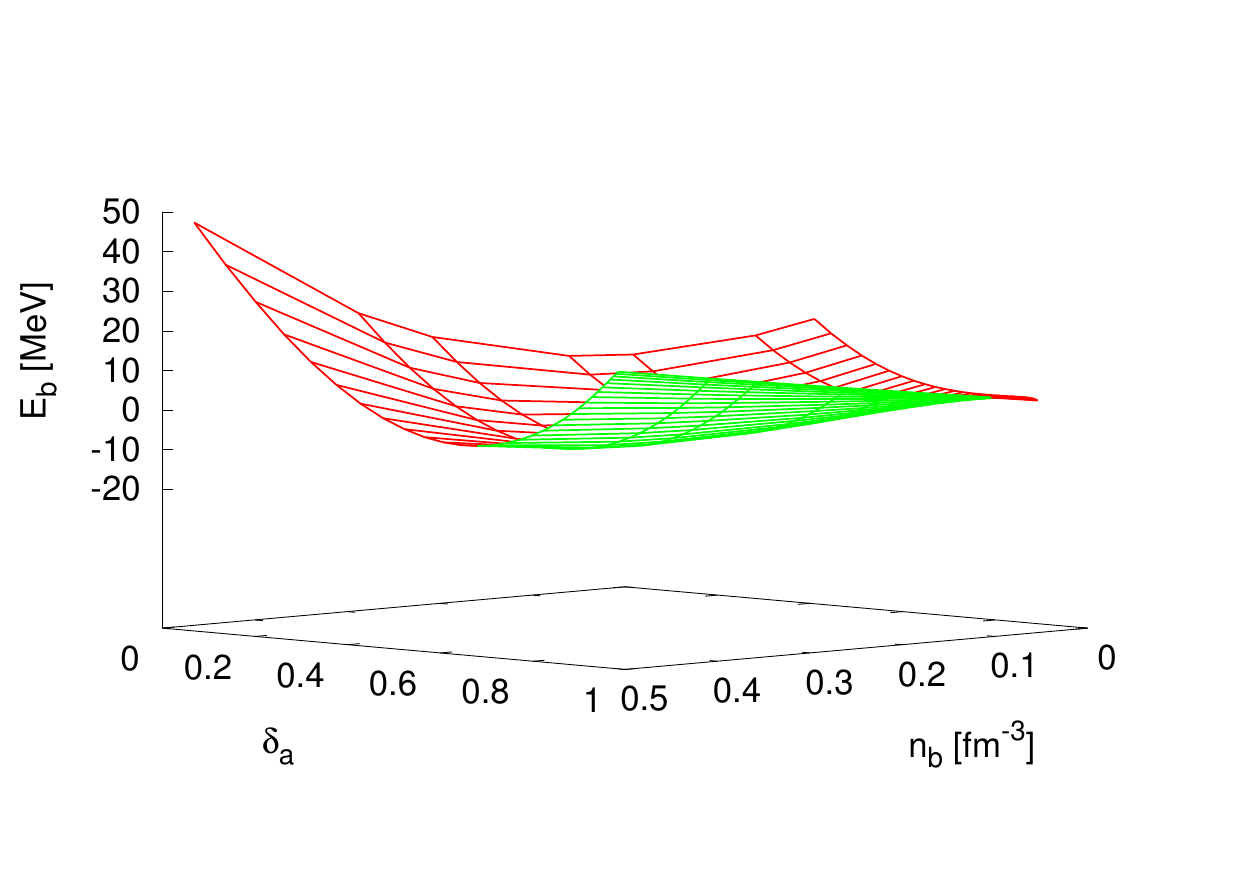} 
\caption{The binding energy of  asymmetric nuclear matter as a function of baryon number density $n_{b}$ and the isospin asymmetry parameter $\delta_{a}$.}
\label{fig:ebin}
\end{figure}

The aim of this paper is to study the high density behaviour of the symmetry energy.
Information on the symmetry energy is encoded in the EoS of asymmetric nuclear matter, so the problem of knowing the density dependence of the symmetry energy  comes down to the analysis\
 of the isospin dependent properties of nuclear matter.  The  exact form of the EoS of asymmetric nuclear matter, calculated for the   TM1 parametrisation is presented in Fig.\ref{fig:ebin}. The EoS is a function of both the isospin asymmetry parameter $\delta_{a}$ and baryon density $n_{b}$. The choice of this parametrization is dictated by the fact that it gives a correspondingly stiff EoS for the matter of a neutron star, which allows for the construction of neutron star models with  the values of masses  in the range acceptable by observations. Additionally, the TM1 parameter set satisfactorily describes  properties of both finite nuclei and nuclear matter. By increasing the value of the parameter $\delta_{a}$,  more asymmetric, less bound matter is obtained. The minimum  of the binding energy for a fixed value of $\delta_{a}$  determines the position of equilibrium density $n_{0}^{a}$, which for increasing asymmetry is shifted towards lower densities. For the case of symmetric nuclear matter, the equality $ n_ {0} ^ {a} = n_ {0} $ holds, where $ n_ {0} $ is the density of saturation.
In the relativistic mean field
approach there is a linear density dependence of the symmetry
energy \cite{1977:Chin, 1994:Sugahara}.
Calculations based on variational many-body theory performed by
Wiringa et al. \cite{1988:Wiringa} with the use of the Argonne two
body potential av14 or Urbana uv14 supplemented with the
three-body potential UVII or TNI lead to the form of the symmetry
energy which initially increases with density and after reaching a
maximum decreases.  In all calculations the kinetic contribution to the
symmetry energy is almost the same. So the reason for the
discrepancy between different model predictions is the
isospin-dependent potential term in the symmetry energy. In the
relativistic mean field theory this potential term is always
repulsive and increases linearly with density.

\begin{figure}
\centering \includegraphics[clip,width=9cm]{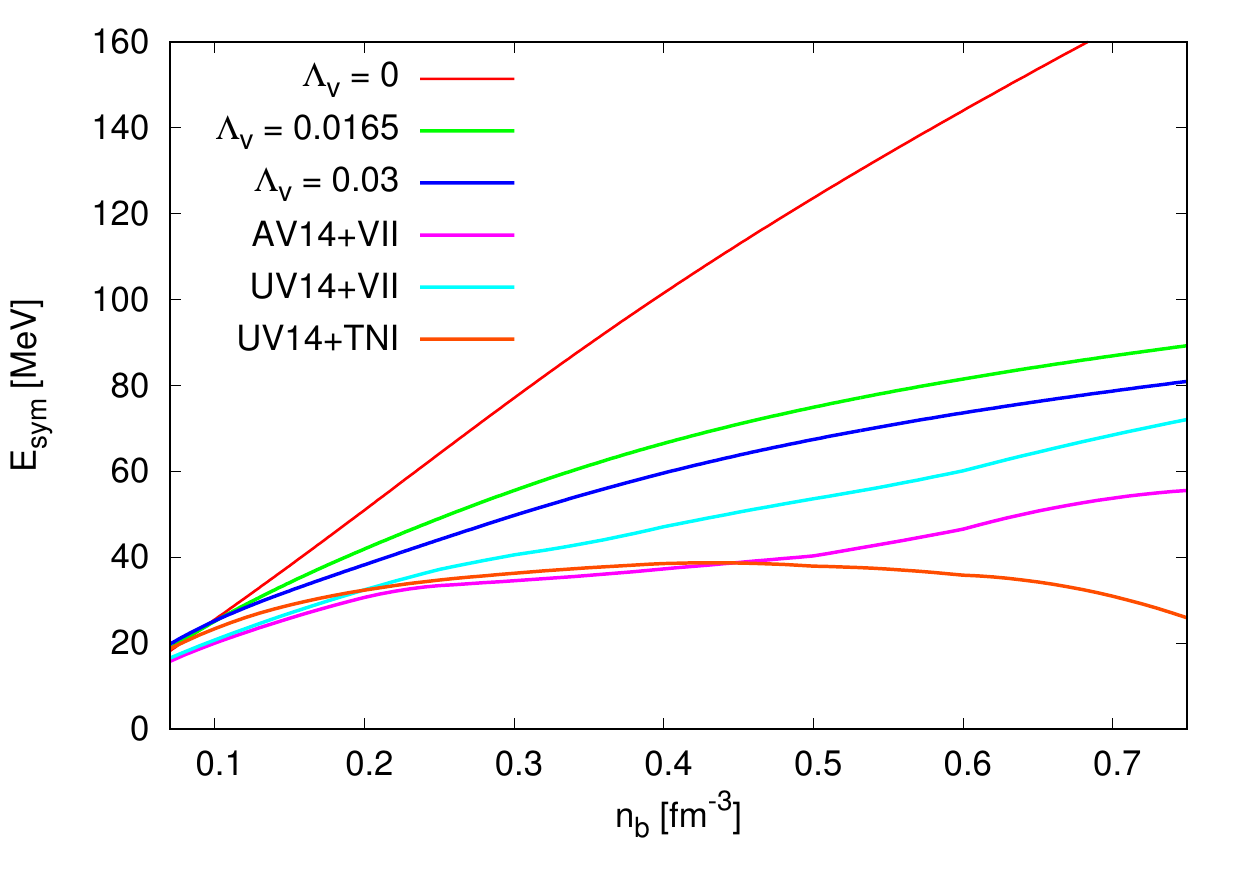} 
\caption{The density dependence of the symmetry energy calculated for different values of the $\Lambda_{V}$ parameter. For comparison, the  results obtained for realistic nuclear potential models are included, cf the text.}
\label{fig:esym_real}
\end{figure}

The density dependence of the symmetry energy calculated for various models is given in Fig.\ref{fig:esym_real}. The parabolic approximation with the use of the standard TM1 parametrization leads to a stiff form of  the symmetry energy.
The extended isovector sector  with  additional $\omega - \rho $ and $\sigma-\rho$ meson couplings  modifies the symmetry energy and allows one to get its softer form.
Results of the analysis presented in Fig.\ref{fig:esym_real} include the cases $\Lambda_{V}=0.0,\, 0.0165\ \text{and}\ 0.03$. The density dependence of the symmetry energy  that approaches solutions obtained for the AV14 and UV14 models, with the Urbana VII (U VII) three nucleon potential, follows from the increasing  value of $\Lambda_{V}$.

\begin{figure}
\centering
\subfigure[]{
\includegraphics[clip,width=7.2cm]{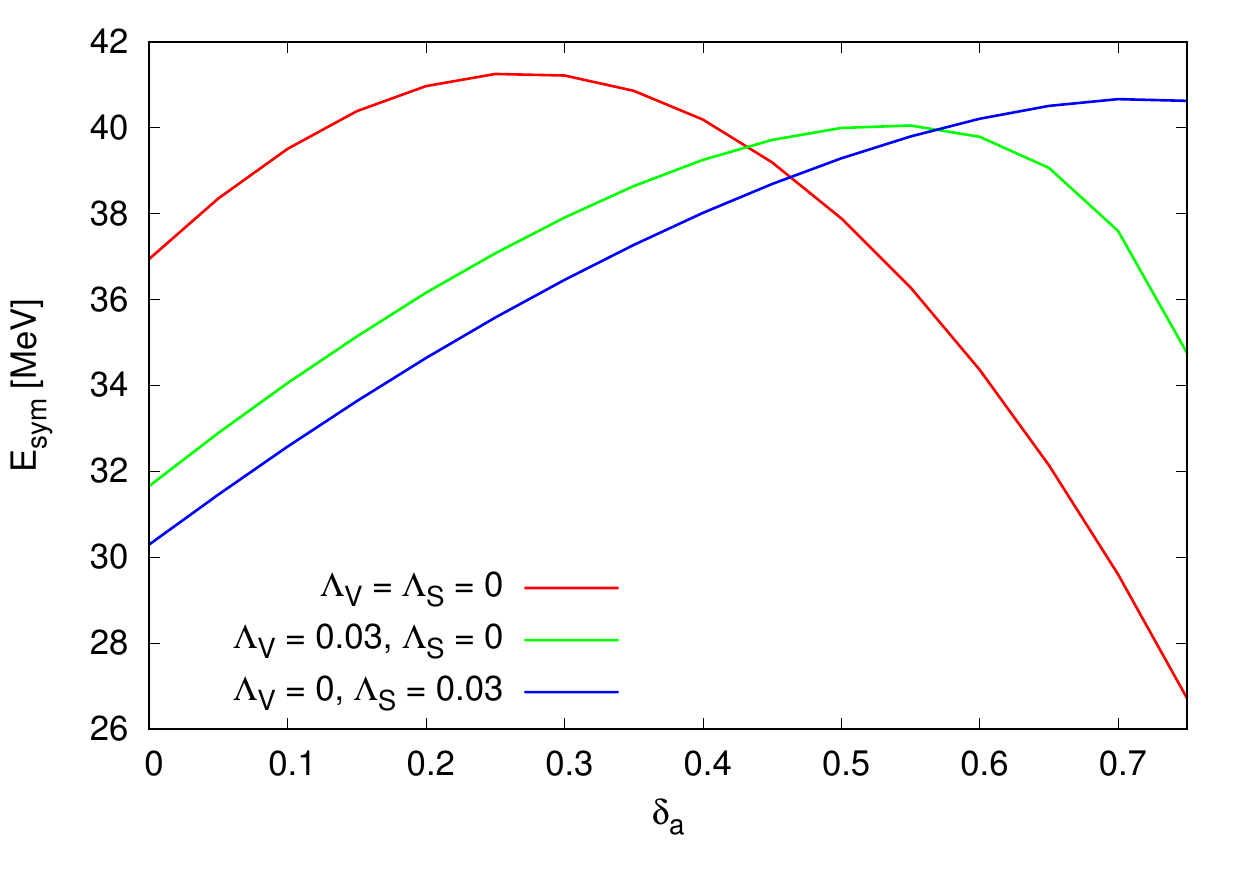} 
}
\subfigure[]{
\includegraphics[clip,width=7.2cm]{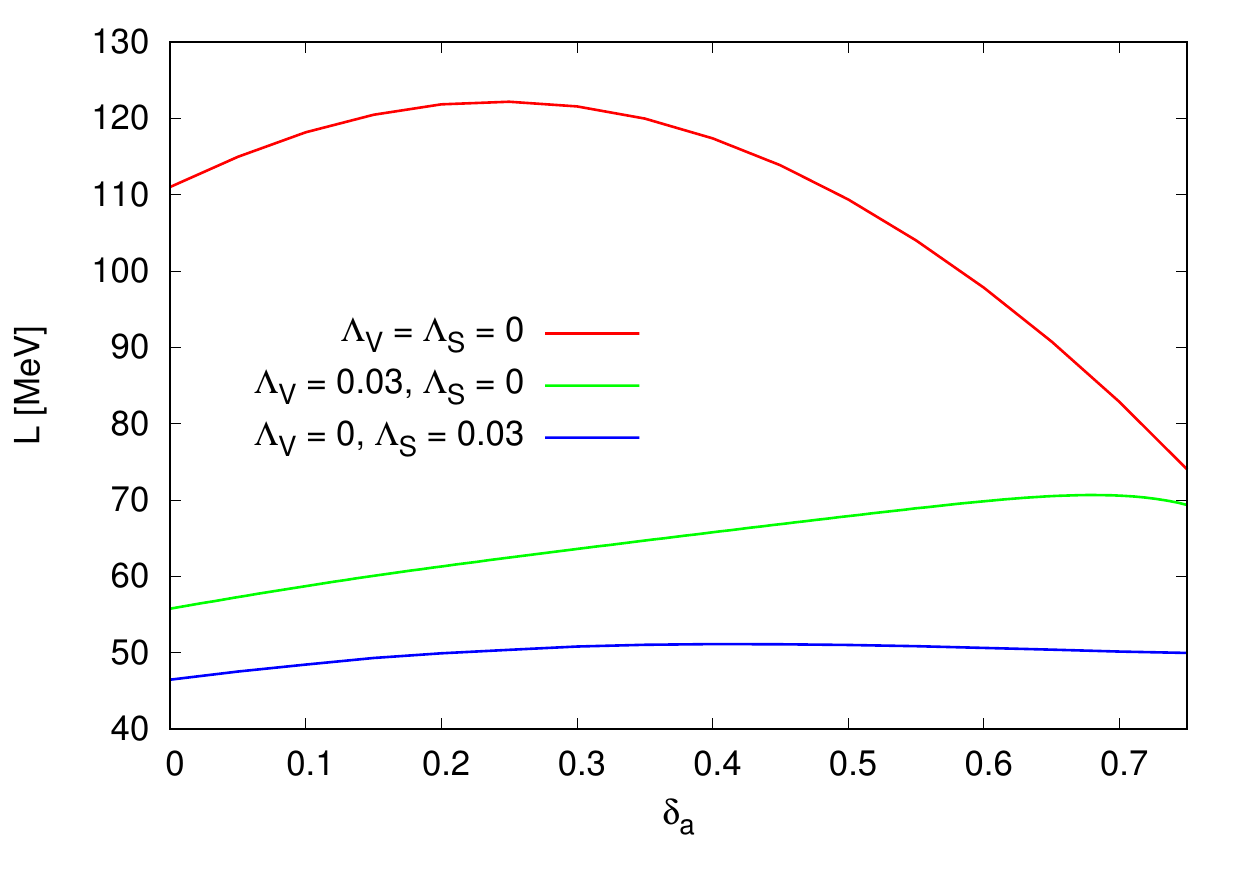} 
}
\caption{The symmetry energy coefficients  calculated at equilibrium density $n_{0}^{a}$ as a function of isospin asymmetry $\delta_{a}$. Left panel shows $E_{2,sym}(n_{0}^{a})$ and the right panel the symmetry energy slope $L(n_{0}^{a})$. Calculations were done for different values of parameters $\Lambda_{V}$  and $\Lambda_{S}$.}
\label{fig:esym_asym}
\end{figure}

In the case of asymmetric nuclear matter, coefficients that characterise  isospin sensitive part of the EoS  depend also on the parameter $\delta_{a}$. Two main parameters, through which the  dependence of the symmetry energy on density can be expressed,  are  the symmetry energy coefficient $E_{sym}(n_{0})$   and the slope parameter $L(n_{0})$, both  given at saturation density $n_{0}$. For asymmetric matter $E_{sym}(n_{0}^{a})$ and $L(n_{0}^{a})$  can be evaluated at equilibrium density $n_{0}^{a}$ and the obtained  coefficients  are changed due to non-zero  $\delta_{a}$. The left panel of Fig.\ref{fig:esym_asym} depicts how the increasing isospin asymmetry
influences the value of the symmetry energy coefficient $E_{sym}(n_{0}^{a})$. Initially, for low asymmetries it increases and  after reaching the maximum, it decreases. Position of the maximum and the range of changes of $E_{sym}(n_{0}^{a})$ depend on  the slope $L$ of the symmetry energy. Isospin dependence of $L(n_{0}^{a})$ reflects the changes in $E_{sym}(n_{0}^{a})$. Inclusion of the  $\omega - \rho$ and $\sigma - \rho$ couplings softens considerably the symmetry energy which leads to a significant reduction of the slope parameter  $L(n_{0}^{a})$. The change of $\Lambda_{V}$ and $\Lambda_{S}$ in the range from 0.0 to 0.03 implies  change in the value of L, from 110.79 MeV to 55.76 MeV for $\Lambda_{V}$ and from 110.79 MeV to 46.47 MeV for $\Lambda_{S}$, which values have been obtained for $\delta_{a}=0$.

\begin{figure}
\centering
\subfigure[]{
\includegraphics[clip,width=7.2cm]{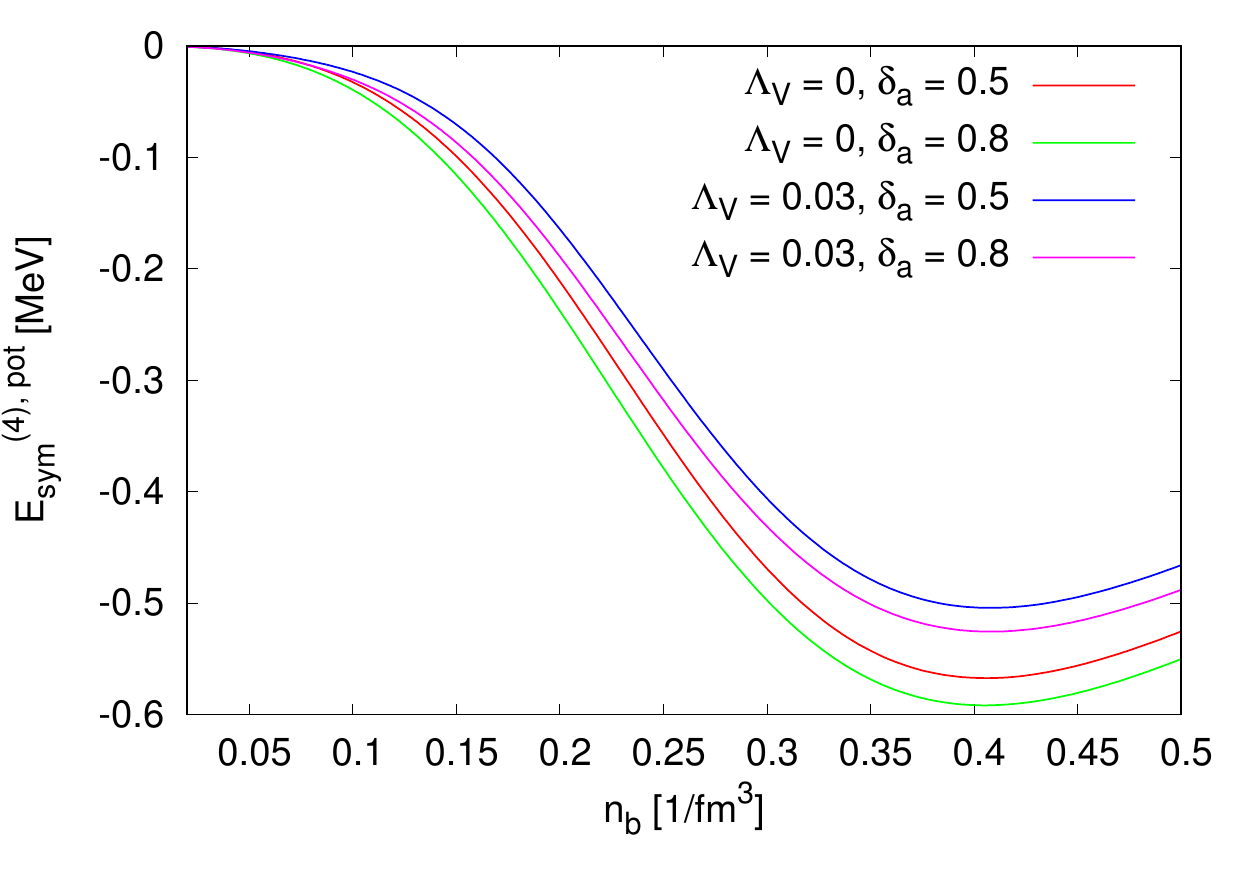} 
}
\subfigure[]{
\includegraphics[clip,width=7.2cm]{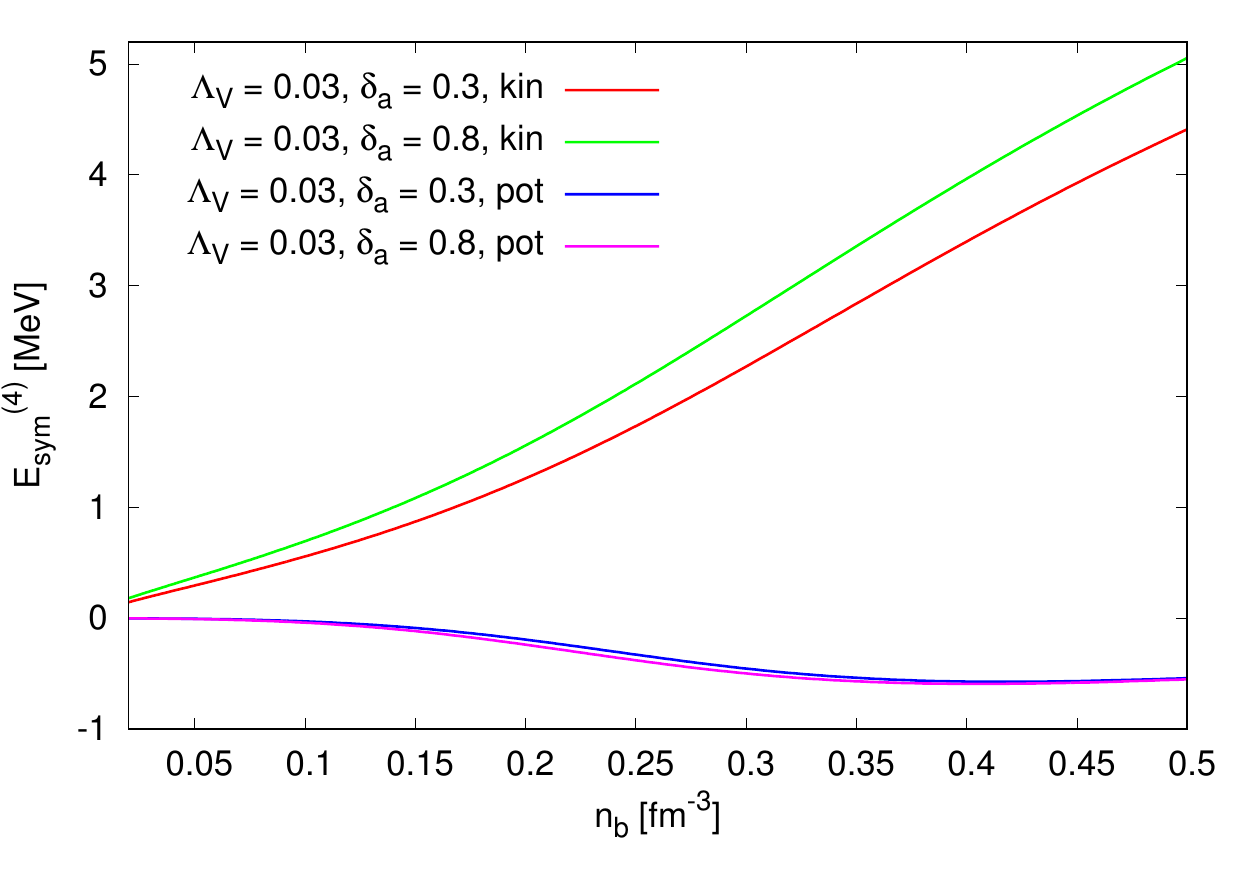} 
}
\caption{The 4-th order symmetry energy terms calculated for TM1 parameterization.  The potential (left panel) and kinetic and potential  contributions (right panel) are presented separately. Calculations of the potential energy have been done for different values of both $\Lambda_{V}$ and $\delta_{a}$ parameters.}
\label{fig:esym4}
\end{figure}

It is expected that the significance of the  4-order term is progressive with increasing density and isospin asymmetry, which are conditions met in neutron star interiors.
In the very general form the symmetry energy may be written as a sum of the kinetic and potential terms, $E_{sym}(n_{b})=E_{s,\, kin}(n_{b})+E_{s,\, pot}(n_{b})$. The kinetic term is due to the shift in neutron and proton Fermi energies whereas the interaction parts includes contributions coming from the isovector sector of the model, which in the simplest case, comprises only  $\rho$ meson. One of possible ways to extend this sector is to consider mixed nonlinear   $\omega-\rho$ and $\sigma-\rho$ couplings.  These non-linear couplings give additional contributions to the potential part of the symmetry energy for both the 2-nd and the 4-th order terms (\ref{eq:esym2}, \ref{eq:esym4k}, \ref{eq:esym4p}). In Fig.\ref{fig:esym4} the results obtained for the 4-th order term and for different value of the isospin asymmetry parameter $\delta_{a}$ are presented.
In the left panel, the effect of both  $\Lambda_{V}$ and $\delta_{a}$  is presented. The nonzero value of $\Lambda_{V}$  leads to higher values of minima for the potential energy. The same effect is obtained for the reduced isospin asymmetry. The right panel of this figure allows one to compare  the kinetic and potential symmetry energy of the 4-th order terms and quantifies  differences in the magnitude between them. Results obtained for the model under consideration show that $E_{4,sym}(n_{b})$ increases with density.  Values of this energy  are approximately between 1 MeV and 4.5 MeV for the density range
$(n_{0},3n_{0})$. In the case of kinetic energy, the influence of  $\Lambda_{V}$  is  small therefore the  results  for only one value $\Lambda_{V}=0.03$ are given.

\begin{figure}
\centering
\includegraphics[clip,width=10cm]{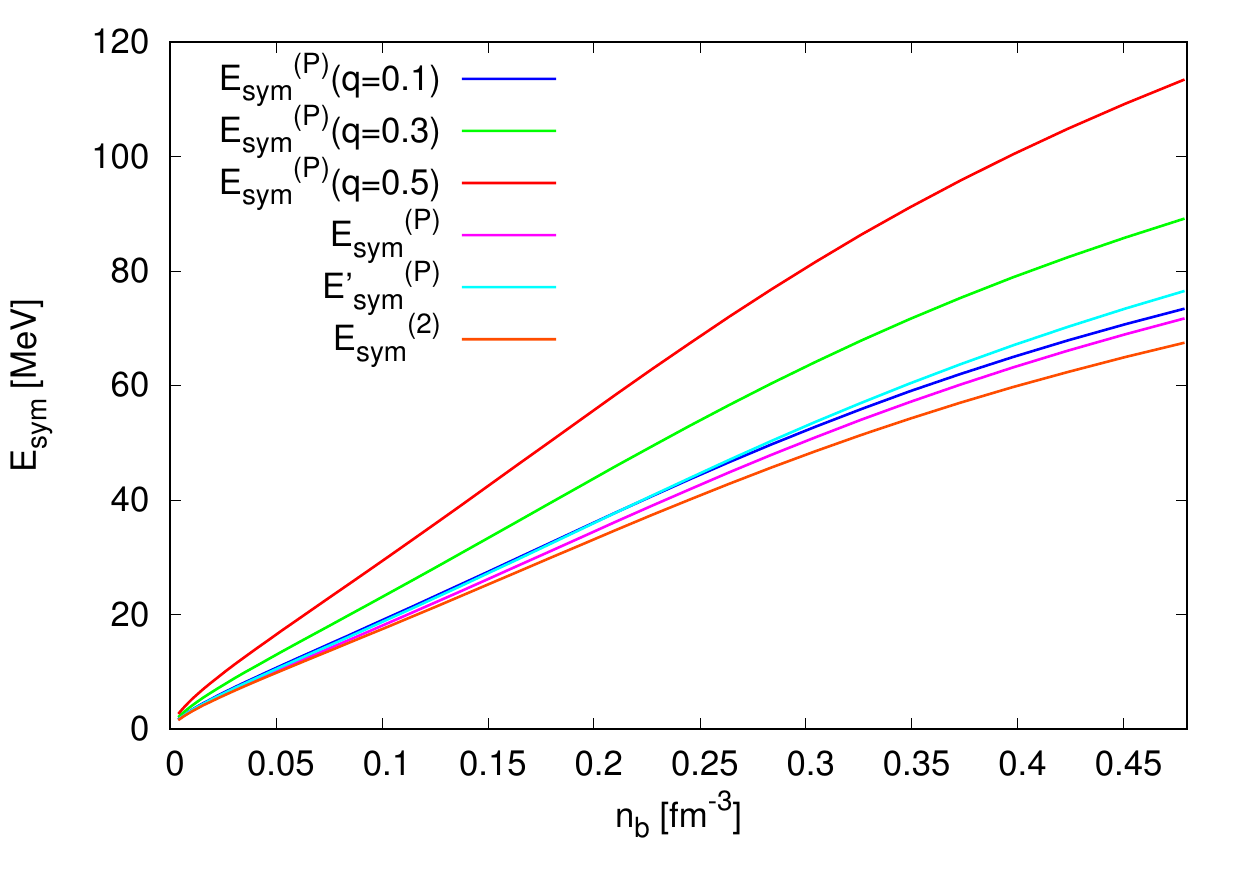} 
\caption{The density dependence of the symmetry energy calculated with the use of Pad{\'e} approximation. The presented  curves starting from the softest one correspond to:  $E^{(P)}_{sym}$ - the Pad{\'e} approximation calculated for the presented model (pink line),   $E^{'(P)}_{sym}$ - the  Pad{\'e} approximation calculated for the  modified model with the increased value of the potential part of the symmetry energy (cyan line), a sequence of curves  obtained for the case of $E_{2,sym}(n_{b})$ and $E_{4,sym}(n_{b})$ self-similar density dependence, for different values of the parameter $q$ (see the text). Results have been compared with the one obtained for parabolic approximation (orange line).}
\label{fig:esymX}
\end{figure}

\begin{figure}
\centering
\subfigure[]{
\includegraphics[clip,width=7.2cm]{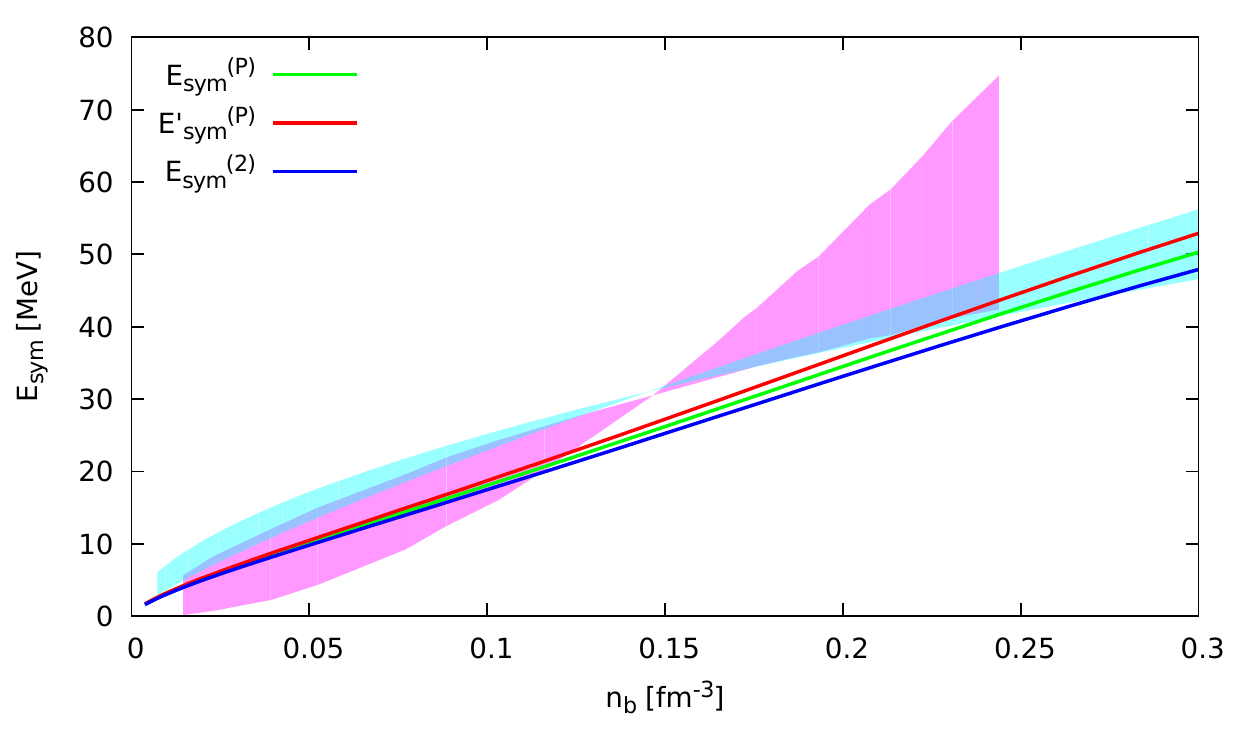} 
}
\subfigure[]{
\includegraphics[clip,width=7.2cm]{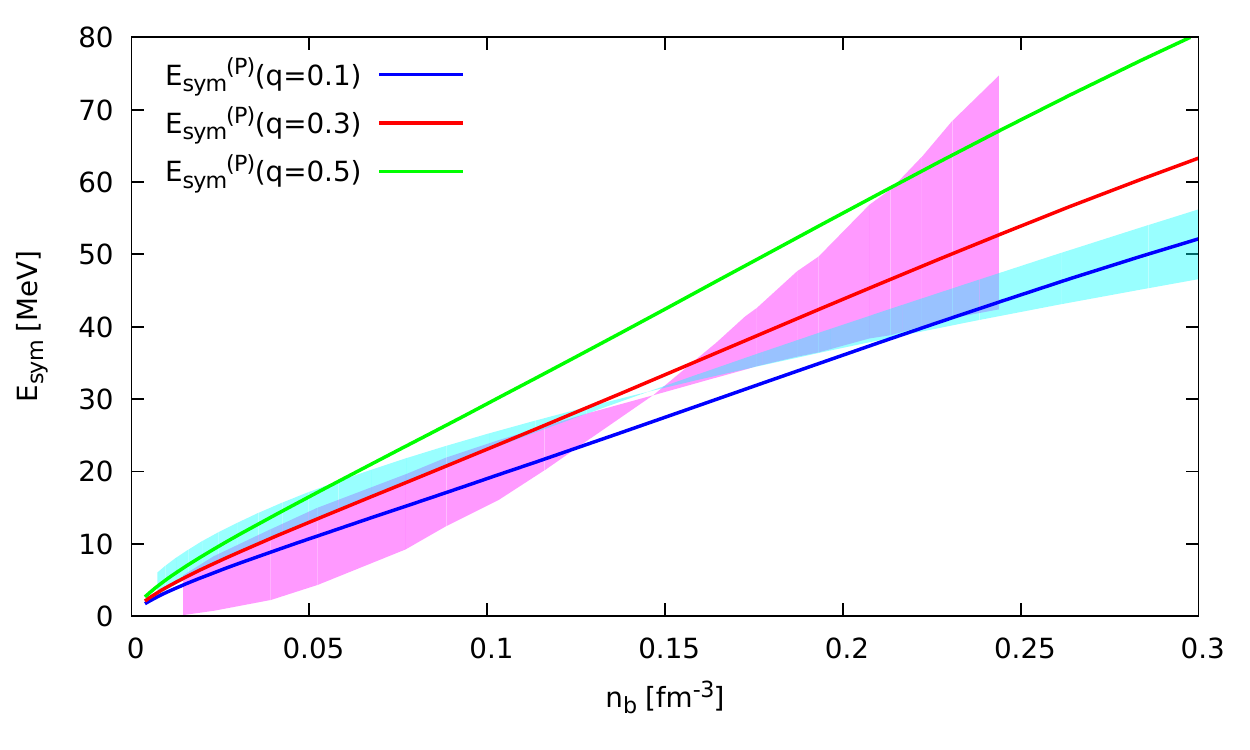} 
}
\caption{Experimental constraints obtained for the density dependence of the symmetry energy. The pink band represents an attempt to create a model-independent limitation of the high-density form of the symmetry energy \cite{2013:Cozma}, whereas the blue band corresponds to the results derived by Russotto et al.\cite{2016:Russotto} }
\label{fig:ensym_exp}
\end{figure}

Because the symmetry energy is studied in systems with high asymmetry and density, therefore  it is necessary to move away from the parabolic approximation. The first step is to take into account the fourth order term in the description of the function, which determines the symmetry energy.
Further analysis  which bases on the Pad{\'e} approximation is presented in the right panel of Fig.\ref{fig:esymX}.
One major advantage of the  Pad{\'e} method lies in the fact that, according to theoretical foundations (Section III), this approximation is a more accurate representation of the function $\varepsilon (n_{b},\delta_{a})$. Additionally, it introduces the explicit dependence on  $\delta_{a}$, which is of particular importance when examining a nuclear matter with a high degree of isospin asymmetry.
Results  presented in  Fig.\ref{fig:esymX} were obtained for $\delta_{a}=0.8$. From this figure one can see that Pad{\'e} approximation leads to a stiffer form of the symmetry energy when compared to the parabolic approximation. These solutions have been achieved for the TM1 parametrization for the model, which gives the slope parameter $L=56$ MeV. In this case the 4-th order term of the Taylor series expansion,  $E_{sym}^{4}$,
has the potential energy, which is only a few percent of the kinetic energy. If the assumption is made that the kinetic and potential energy terms are of comparable magnitude  then    further stiffening of the symmetry energy can be obtained.
In the case of the Pad{\' e} approximation, the symmetry energy given by formula (\ref{eq:esym_Pade}) depends on the value of isospin asymmetry $\delta_{a}$ and the ratio $E_{4,sym}(n_{b})/E_{2,sym}(n_{b})$.
 The available experimental data provide little information about the relationship of $E_{4,sym}(n_{b})$ and $E_{2,sym}(n_{b})$. Certainly,  the ratio $E_{4,sym}(n_{b})/E_{2,sym}(n_{b})$ treated as a function of $n_{b}$ is of bounded variation. One can even risk saying that $E_{4,sym}(n_{b})$ is practically proportional to $E_{2,sym}(n_{b})$ in the available energy range. Such an  assumption leads to interesting consequences. The theory of ordinary differential equations \cite{1983:Kamke} says that for  given bondary values one gets an infinite series $q_n$ of  possible values  of $q=E_{4,sym}(n_{b})/E_{2,sym}(n_{b})$ that can be compared with experimental data. A detailed analysis of this case is in preparation. It is possible to estimate constrains on the value of $E_{4,sym}(n_{b})$ derived for different theoretical approaches and on the basis of experimental data (Table \ref{tab:S4_theory}). It varies from a few to several dozen percent. By introducing a parameter which describes the ratio, say  $q=E_{4,sym}(n_{b})/E_{2,sym}(n_{b})$,  it is possible to analyse how  the value of this parameter  influences the final form of symmetry energy.
Results are depicted in   Fig.\ref{fig:esymX}. Increasing the value of the $E_{4,sym}(n_{b})$, which corresponds to a higher value of the parameter $q$, stiffens the form the symmetry energy. More precise functional form of the ratio $E_{4,sym}(n_{b})/E_{2,sym}(n_{b})$ is beyond the scope of this analysis.
In this figure solutions obtained for various forms of the symmetry energy are included. The main part of the results concerns the Pad{\'e} approximation. In this approach the solution obtained for the considered model ($E_{sym}^{(P)}$) is represented by the cyan  curve, whereas the pink line ($E_{sym}^{'(P)}$) is for the same model yet with the increased value of potential part of the symmetry energy.  These  solutions have been obtained for $\Lambda_{V}=0.03$, and for very high isospin asymmetry $\delta_{a} =0.8$ and compared with the results derived for parabolic approximation. The latter case gives the softens symmetry energy.
The resulting solutions for different values of $q$ are represented as a series of models for which  the assumption regarding the approximate proportionality of $E_{4,sym}(n_{b})$ and $E_{2,sym}(n_{b})$ was adopted. The highest value of $q$ leads to the stiffest form of the symmetry energy.
In the next step, various  modification methods of the symmetry energy were compared with experimental limitations.
A summary of  theoretical results and the corresponding experimental constraints are presented in  Fig.\ref{fig:ensym_exp}. The band shaded in blue represents experimental constraints obtained by Russotto et al. \cite{2016:Russotto} whereas, the pink one corresponds to solutions that were given in the paper by Cozma et al. \cite{2013:Cozma}.
The latter results were obtained as an attempt to create a model-independent limitation of the high-density form of the symmetry energy.
The presented constraints differ in their approach to the isovector part of the EoS and the transport model involved. The model-independent solution uses the Gogny inspired (momentum-dependent) potential as well the Tubingen QMD model \cite{2011:Russotto}, while the results obtained by Russoto et al. are based  on the power law parametrization of the symmetry energy and the UrQMD transport model \cite{2016:Russotto}. In the left panel of Fig.\ref{fig:ensym_exp}, the presented experimental limitations were compared with the results obtained for the Pad{\'e} approach. It is evident that solutions derived on the basis of the considered model, for the  high value of the isospin asymmetry are only marginally  consistent with the high density limit of the presented constraints.  In the case of the Pad{\'e} approximation and assuming proportionality of $E_{4,sym}(n_{b})$ and $E_{2,sym}(n_{b})$ satisfactory compliance with experimental limitations can be obtained.
\section{Possible astrophysical implications}
\begin{figure}
\centering
\includegraphics[clip,width=9cm]{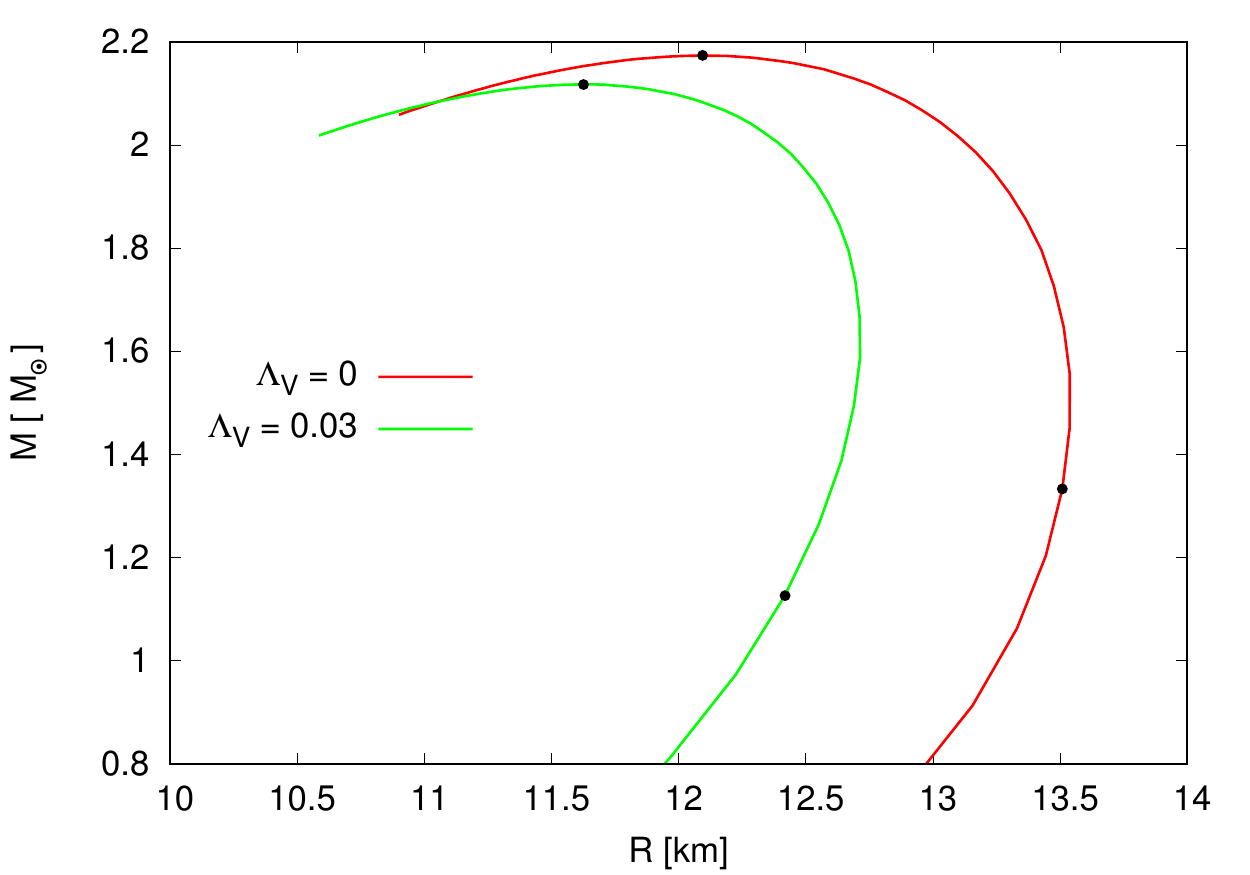} 
\caption{The mass-radius relations obtained for the TM1 parametrisation for $\Lambda_{V}=0$ and for $\Lambda_{V}=0.03$. Dots represent the selected stellar configurations: the maximum mass configuration and the one caracterised by the central density value equal to $2 n_0$.}
\label{fig:RM}
\end{figure}

\begin{figure}
\centering
\subfigure[]{
\includegraphics[clip,width=7.2cm]{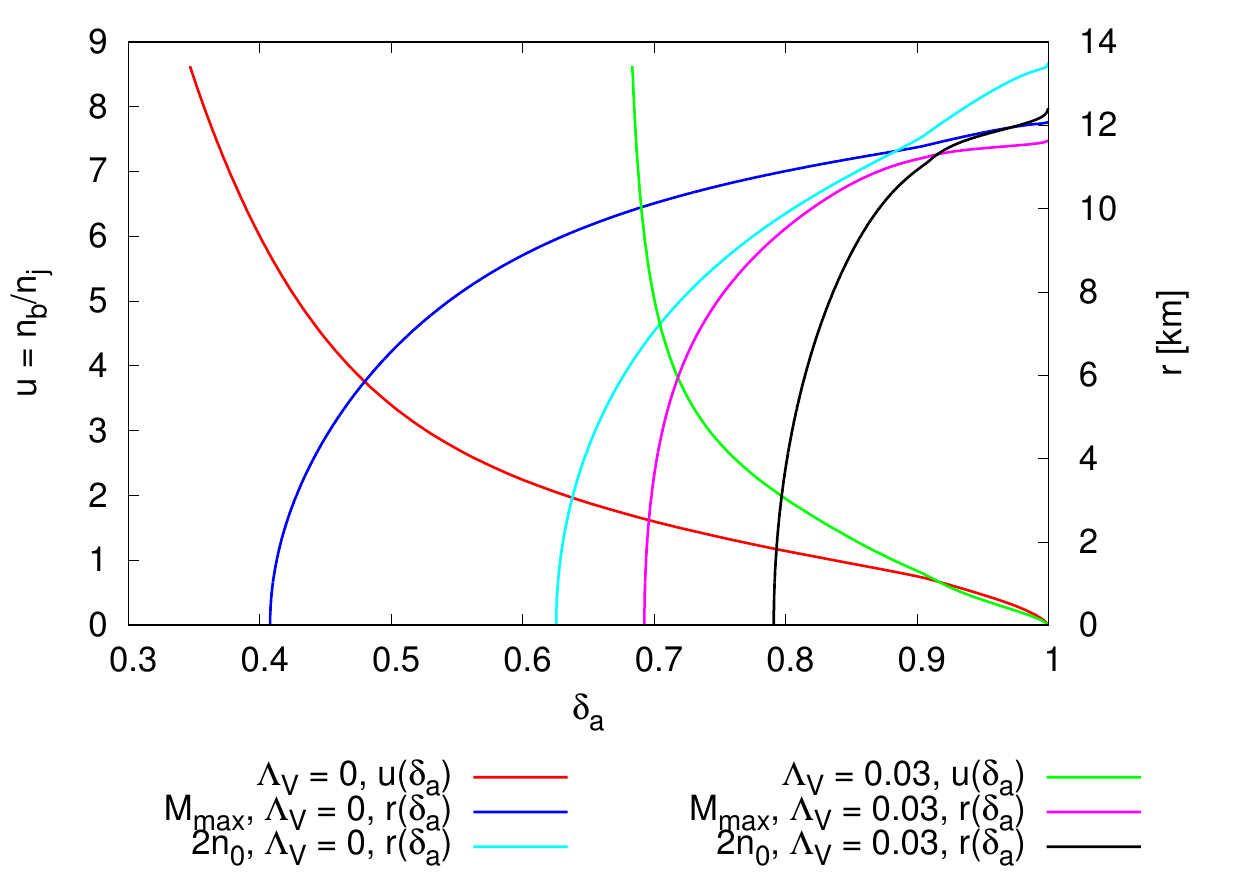} 
}
\subfigure[]{
\includegraphics[clip,width=7.2cm]{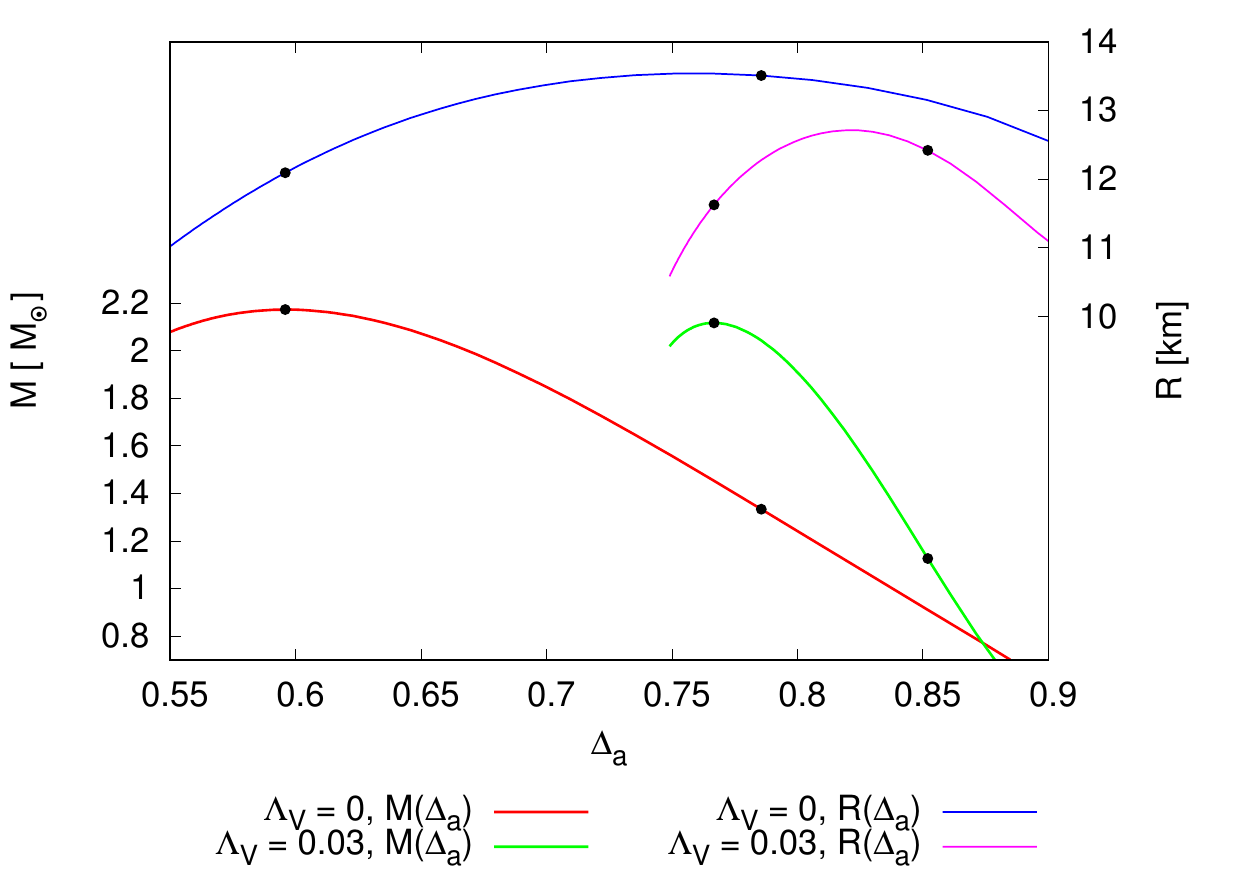} 
}
\caption{In the left panel the density and radial dependence of the isospin asymmetry parameter is plotted. Results have been obtained for $\Lambda_{V}=0.0$ and $0.03$. In the case of radial dependence for each value of $\Lambda_V$ the solutions for two stellar configurations are included. These are the maximum mass configuration and the configuration with $\rho_{c}=2n_{0}$. The right panel shows relations $M(\Delta_{a})$ and $R(\Delta_{a})$, where $M$ and $R$ are the total mass and radius of neutron stars calculated for TM1 parametrisation with $\Lambda_{V}=0.0$ and $0.03$. $\Delta_{a}$ denotes the total relative neutron excess obtained for a given stellar configuration. Positions of the selected configurations are marked by dots.}
\label{fig:FA}
\end{figure}

\begin{figure}
\centering \includegraphics[clip,width=9cm]{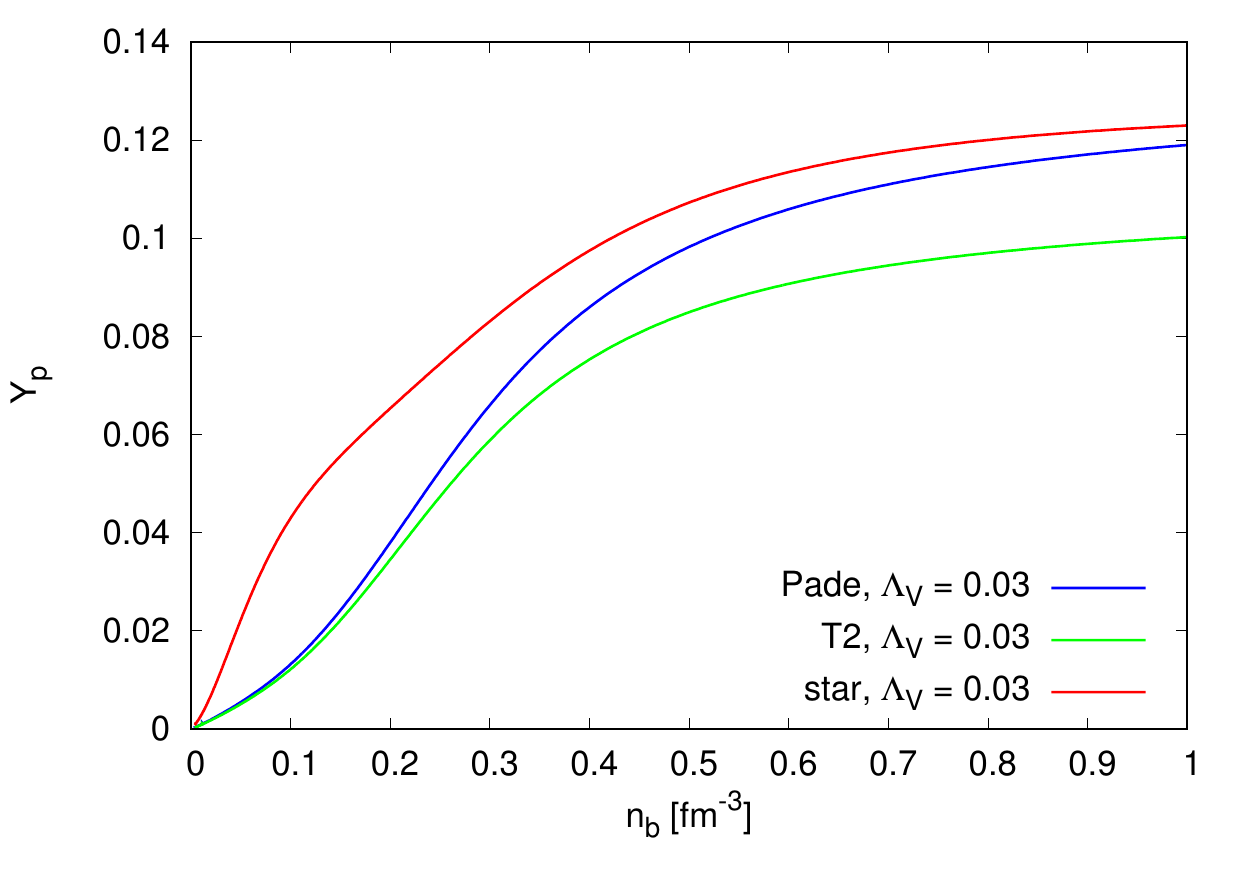} 
\caption{The density dependence of the relative proton fraction $Y_{p}$ for the neutron star matter obtained for $\Lambda_{V}=0.03$. For comparison
$Y_{p}(n_b)$ calculated for the symmetry energy in the parabolic  and in  Pad{\'e} approximations are also given.}
\label{fig:Yp}
\end{figure}

The analysis of  asymmetric nuclear matter is of great importance in astrophysical applications, especially regarding  the modelling of neutron star interiors. This allows one to study how the change in isospin asymmetry translates into a change in the properties of neutron stars.
Their parameters such as the mass and radius  can be calculated basing on the equation of hydrostatic equilibrium - the Tolman-Oppenheimer-Volkoff (TOV) equations:
\begin{eqnarray}\label{TOV}
\frac{dP(r)}{dr}&=&\frac{-G\left(\epsilon (r) +  P(r)\right)\left(m(r)+4\pi r^3P(r)\right)}{r^{2}\left(1-\frac{2Gm(r)}{r}\right)}\\ \nonumber
\frac{dm(r)}{dr}&=&4\pi r^{2} \epsilon (r) \\ \nonumber
\frac{dn_{b}(r)}{dr}&=&4\pi r^{2}\left(1-\frac{2Gm(r)}{r}\right)^{-1/2},
\end{eqnarray}
where $m$ and $n_b$ denote the enclosed gravitational mass and baryon number, respectively.  Supplied with the EoS, which have been derived
for the TM1 parametrisation, for  $\Lambda_{V}=0$ and $\Lambda_{V}=0.03$ the mass and radius of a neutron star for a given central density $\rho_{c}$ can be calculated. Solutions of the TOV equation  in the form of R-M diagrams are presented in  Fig.\ref{fig:RM}.
  For the analysis, two characteristic stellar configurations were selected. The one for which the central density $\rho_{c}$ is of the order of $2 n_{0}$ and the second being the maximum mass configuration.
The choice of the first configuration was dictated by the desire to examine the interior of a star, whose central density is comparable to the density value achieved in the HIC experiments.
Analysis of the selected isospin sensitive observables allows for the estimation of this density. Its  upper limit  is within the range $(2n_{0},3n_{0})$.
  The chosen stellar configurations are marked by dots on the R-M diagrams. The form of the obtained R-M diagrams confirms the conclusion about  the role of the parameter $\Lambda_{V}$, whose presence has a small effect on the value of the maximum mass, yet significantly changes the value of the stellar radius.
Having obtaind the solution of TOV equations, it is possible to study the internal structure of a neutron star.
In this paper  the analysis focuses on  the isospin asymmetry parameter $\delta_{a}$. This type of studies was carried out in various aspects. The left panel of Fig.\ref{fig:FA} shows the density and radial dependences of $\delta_{a}$. The general conclusion that can be drawn is that the
presence of $ \omega - \rho$ coupling leads to more asymmetric  matter. This also can be seen by comparing radial dependence of $\delta_{a}$ calculated for  selected configurations.
Instead of $\delta_{a}$, the stellar radius and mass can be plotted against the parameter $\Delta_{a}$, which  determines the total relative excess of neutrons in a star. It was calculated in the same way as  $\delta_{a}$ however, for $N_{n}$, and $N_{p}$ defined as $ N_{i}=  \int_{0}^{R}\frac{n_{i}(r)}{\sqrt{1-\frac{2m(r)}{r}}}dr$,
where $m(r)$ is the mass enclosed inside the radius $r$ and $n_{i}, (i=n, p) $ denotes  number density of neutrons and protons. The relation between the mass and radius of a star and the quantity $\Delta_{a}$ is plotted in the right panel of Fig.\ref{fig:FA}.
The analysis of the $M(\Delta_{a})$ dependence shows that the increase in the value of parameter $\Lambda_{V}$ causes a change in the range of $\Delta_{a}$ values that can be achieved up to the $M_{max}$ configuration. For $\Lambda_{V}=0.0,$ $\Delta_{a}$ for $M_{max}$ configuration is of the order of 0.6, whereas for $\Lambda_{V}=0.03,$ $\Delta_{a}$ for $M_{max}$ equals $\approx 0.77$. The extremely large value of $\Delta_{a} \approx 0.86$ was achieved for the configuration characterised by $\Lambda_{V}=0.03$ and  $\rho_{c}= 2n_{0}$. The remaining upper curves in this figure show the dependence $R(\Delta_{a})$.
Results containing values of parameters that characterize individual stellar configurations are gathered in the Table \ref{tab:prop}.
\begin{table}

\caption{Parameters of the selected neutron star configurations} \label{tab:prop}
\begin{center}
\begin{tabular}{l|l|l|l|l|l|l}
\hline
&$\Lambda_{V}$&$\Delta_{a}$& $\rho_{c}/n_{0}$ & M[M$_{\odot}$] &R [km]& $\delta_{a}(0)$\\ \hline
M$_{max}$& 0.03&0.77&6.05&2.11&11.6&0.69\\ \hline
$2\times n_{0}$&0.03&0.85&2.08&1.13&12.42&0.79\\ \hline
M$_{max}$&0.0&0.6&5.71&2.17&12.09&0.41\\ \hline
$2\times n_{0}$&0.0&0.79&2.05&1.33&13.05&0.63\\ \hline
\end{tabular}
\end{center}
\end{table}
 The symmetry energy is decisive for the proton fraction $Y_{p}=n_{p}/n_{b}$ in $\beta$-stable neutron star matter, being one of the most isospin sensitive observable. Determining its value is based on
the problem of  chemical composition of a neutron  star, which  has been discussed in Section VI A.  In the case of cold neutrino free matter $\mu_{asym}=\mu_{e}$.
This can be related to the symmetry energy through the equation $\mu_{asym}=4\delta_{a}E_{2sym}(n_b)$. This relation holds when only  the 2-nd order term of the Taylor expansion is included. Taking into account additionally the 4-th order term the following result can be derived $\mu_{asym}=4\delta_{a}E_{2sym}(n_b) + 8 \delta_{a}^{3}E_{4sym}$. In Fig.\ref{fig:Yp} relative concentrations of protons obtained for the considered model have been presented. The results obtained refer to the neutron star nucleus
The red, upper curve represents density dependence of $Y_{p}$ that follows directly from the considered model of the neutron star, calculated for $\Lambda_{V}=0.03$. The remaining curves show solutions obtained for the symmetry energy in the parabolic approximation $T2,\ \Lambda_{V}=0.03$, and the Pad{\'e} approximation. In the latter case the result that much better reproduces the actual concentration of protons obtained for a specific neutron star matter model is derived.
\section{Conclusions}

The purpose of this paper was to analyse the isospin dependent properties of asymmetric nuclear matter and the involved definition of the symmetry energy. The standard definition of the symmetry energy as the second and fourth order term in Taylor expansion is compared with the one obtained in the  Pad\'{e} approximation. One should expect that the results are not identical due to the different convergence properties of approximations. The main difference  consists in  the fact that the  symmetry energy in the Pad\'{e} approximation itself becomes a  function of  both the baryon density and isospin asymmetry.  Switching on the explicit dependence on $\delta_a$ reveals itself in highly isospin asymmetric systems. The virtue of this approach is that one can trace this  dependence. The performed analysis indicates that the  important  factor influencing the form of the symmetry energy is  the ratio $E_{4, sym}(n_b)/ E_{2, sym}(n_b).$  This reveals one of the main sources of uncertainties as the functional form of $E_{4, sym}(n_b)$ is practically unknown. In this investigation, only the case when this ratio is approximately constant was considered and compared with experimental constraints. Going beyond this approximation and accepting a more involved functional form of the ratio $E_{4, sym}(n_b)/ E_{2, sym}(n_b)$ opens new interesting possibilities of further studies. The performed analysis of the equilibrium proton fraction in neutron star matter demonstrates the advantage of the Pad\'{e} approximation approach.

\end{document}